\documentclass[american]{article}
\usepackage[T1]{fontenc}
\usepackage[utf8]{inputenc}
\usepackage{geometry}
\usepackage{color}
\usepackage{xcolor}
\usepackage{array}
\usepackage{calc}
\usepackage{amsmath}
\usepackage{amssymb}
\usepackage{graphicx}
\usepackage{esint}
\usepackage{caption}
\usepackage{subcaption}
\usepackage{xspace}
\usepackage{lineno}

\definecolor{citecolor}{rgb}{0.,0.5,0.1}
\usepackage[hypertexnames,setpagesize,%
    pdftex,%
    colorlinks,%
    citecolor=citecolor,%
    linkcolor=blue,
    hyperindex,%
    plainpages=false,%
    bookmarksopen,%
    bookmarksnumbered%
  ]{hyperref}
  
\usepackage[numbers,sort&compress]{natbib}

\newcommand\myref{\refstepcounter{equation}\theequation}

\begin{document}

\title{
Recursion relations for multi-gluon off-shell amplitudes
on the light-front  and  Wilson lines
}

\date{\vspace{-3ex}}

\author{C.~Cruz-Santiago$^{(a)}$, P.~Kotko$^{(a)}$, A.M.~Sta\'sto$^{(a,b)}$ \vspace*{0.2cm}\\
$^{(a)}$  {\it \small  Department of Physics, 
 The Pennsylvania State University, University Park, PA 16802, United States}\\
$^{(b)}$ \it \small H. Niewodnicza\'nski Institute of Nuclear Physics, Polish Academy of Sciences, Krak\'ow, Poland}
\maketitle

\begin{abstract}
We analyze the off-shell scattering amplitudes in the framework of the light-front perturbation theory.
It is shown that the previously derived recursion relation between 
tree level
off-shell amplitudes in this formalism 
actually resums whole classes of graphs into a Wilson line.
More precisely, we establish a correspondence between the light-front methods for the computation of the off-shell amplitudes and the approach which makes use of the matrix elements of straight infinite Wilson lines, which are
manifestly gauge invariant objects.
Furthermore, since it is needed to explicitly verify the gauge invariance of light-front amplitudes, it is demonstrated that the Ward identities in this framework need  additional instantaneous terms in the light-front graphs. 
\end{abstract}

\section{Introduction}

Scattering amplitudes in quantum chromodynamics (QCD) with on-shell initial and final states 
are basic objects which can be calculated using perturbative methods
of quantum field theory. Together with the suitable factorization
theorems \cite{Collins:1985ue,Collins:2011zzd} and parametrizations of the non-perturbative parton densities \cite{Martin:2009iq,Lai:2010vv}
and fragmentation functions \cite{Kniehl:2000fe} they are used to evaluate cross sections
for various observables at high transverse momenta in the processes
that occur in high energy collisions. Over the past two decades there has been an enormous progress
in the computational techniques of the scattering amplitudes and their implementation in the computer codes for calculating various processes, see
for example \cite{Dixon:1996wi,Mangano:2002ea,Gleisberg:2008fv,Dixon:2010ik,Kleiss:2010hy}. Of particular importance are the calculations
of the on-shell scattering amplitudes with the fixed helicities \cite{Parke:1986gb}, for
a review see \cite{Mangano:1990by,Dixon:1996wi}. Since the amplitudes with different helicity
configurations do not interfere with each other, they can be added
incoherently. On-shell helicity scattering amplitudes can be efficiently
computed using the Berends-Giele recursion methods  \cite{Berends:1987me} which use off-shell currents or the
Britto-Cachazo-Feng-Witten (BCFW) recursion relations \cite{Britto:2004ap,Britto:2005fq} which utilize gauge invariant on-shell amplitudes with shifted complex momenta.

On-shell scattering amplitudes have however some limitations, since
in reality the quarks and gluons are never on-shell particles, and thus are never observed as free states in the experiments. The off-shell matrix elements are more general objects which can be used for the construction of the on-shell scattering amplitudes, like for  example in the above-mentioned Berends-Giele recursion. Furthermore, 
the use of the off-shell matrix elements in the phenomenology together with the unintegrated
parton densities  and appropriate $k_T$ factorization 
approaches
 is the alternative method for the computation of
the cross sections, see for example \cite{Catani:1990xk,Catani:1990eg,Collins:1991ty}. This approach, albeit more theoretically challenging, has the benefit of   taking into account  kinematics more accurately.
This can be essential for example, when computing more exclusive processes which
do require information about the details of the kinematics. One complication
though in using off-shell matrix elements is the condition of the
gauge invariance. Recently, a progress has been made  \cite{vanHameren:2012if,Kotko:2014aba,vanHameren:2014iua}
in the construction of the  off-shell amplitudes which do satisfy Ward identities
and hence obey gauge invariance. The general method \cite{Kotko:2014aba} utilizes infinite Wilson line operators
corresponding to the off-shell gluons, whose directions are defined
by the polarization vectors perpendicular to the momenta of the off-shell
gluons. It has been shown that such a definition of the off-shell matrix
elements satisfies the corresponding Ward identities with respect to the remaining on-shell states and, as such, is
gauge invariant.

In previous works \cite{Motyka:2009gi,CruzSantiago:2013uk,CruzSantiago:2013wz,Cruz-Santiago:2013vta} gluon wave functions, fragmentation functions
and scattering amplitudes for selected helicity configurations have
been derived using the methods of perturbation theory on the light-front \cite{Kogut:1969xa,Lepage:1980fj,Brodsky:1997de}.
In particular, certain interesting recursion relations have been proved between
the off-shell amplitudes, which enabled in turn to construct all tree level
Maximally Helicity Violating (MHV) on-shell amplitudes in this framework \cite{Cruz-Santiago:2013vta}. 

In this paper we shall explain in detail the
physical origin of this recursion relation. Namely, we shall show that
it is a direct consequence of the gauge invariance, as for any off-shell amplitude one can construct
its gauge invariant extension, using for example Wilson lines as in \cite{Kotko:2014aba}. 
Those Wilson lines encode certain recursion, which turns out to be identical to the one obtained within 
the light-front perturbation theory (LFPT). 
Moreover, since we are interested in the gauge invariance properties of the amplitudes within this formulation of QCD, we need a method to check the Ward identities explicitly on the light-front. To this aim, we shall demonstrate that one needs to redefine the rules for the
LFPT in the context of the Ward identities. Namely, in the calculation of the ordinary
light-front diagrams, the minus components of the momenta only occur
in the energy denominators and as such are not conserved. However,
in order to show the Ward identity in the light-front theory the minus
components actually appear in the numerators of the expression for
the amplitudes, because of the replacement of the polarization vector with the momentum.
This results in the additional instantaneous terms which need to be 
considered  in addition to the standard light-front diagrams and which, in fact, restore the full momentum conservation in the vertices, for all light cone components. After this is taken into account, the Ward
identities are satisfied on the light-front for gauge invariant objects as expected.

The outline
of the paper is as follows: in the next section we  introduce
the notation and conventions used throughout the paper, in Sec.~\ref{sec:RecRelation}
we  recall the recursion relation for the off-shell amplitudes on the light-front
which was derived in \cite{Cruz-Santiago:2013vta}. This recursion relation was derived starting from the Berends-Giele like relations for the light-front off-shell amplitudes. We  also recall an expression for the 
off-shell amplitude with $(+\rightarrow+\dots +)$ helicity configuration which  was derived in the light-front.
This amplitude is non-zero for off-shell states and vanishes in the on-shell limit.
  In Sec.~\ref{sec:WardId} we  discuss the Ward identities within the light-front formalism. We shall enforce ourselves with an explicit example of the lowest nontrivial order amplitude $(+\rightarrow -++)$ on the light-front. Using this low-order example we demonstrate that the light-front recursion relation for off-shell amplitude indeed contains a natural and gauge invariant object, which in turn originates in a straight infinite Wilson line. 
In Sec.~\ref{sec:GI} we
 generalize this picture to the scattering amplitudes with arbitrary large number of external gluons.
Finally
in Sec.~\ref{sec:Conclusions} we state our conclusions. Appendices  contain  useful
identities and  technical details regarding some formulae discussed in the main text.

\section{Notation and conventions}

\label{sec:Notation}

The decomposition of any four vector $u$ in the light cone basis reads
\begin{equation}
u^{\mu}=\frac{1}{2}u^{+}\eta^{\mu}+\frac{1}{2}u^{-}\tilde{\eta}^{\mu}+u_{T}^{\mu}\, ,\label{eq:LCdecomp}
\end{equation}
with $u_{\perp}^{\mu}=\left(0,u^{1},u^{2},0\right)\equiv\left(0,\vec{u}_{\perp},0\right)$
and the minus and plus components are defined by a projection
on the following null four vectors
\begin{gather}
\eta=\left(1,0,0,1\right),\label{eq:eta}\\
\tilde{\eta}=\left(1,0,0,-1\right).\label{eq:etatild}
\end{gather}
A scalar product in this basis can be written as $u\cdot w=\frac{1}{2}u^{+}w^{-}+\frac{1}{2}u^{-}w^{+}-\vec{u}_{\perp}\cdot\vec{w}_{\perp}$.

We shall be dealing with  gluon amplitudes with definite helicities throughout this
paper. A polarization vector $\varepsilon_{i}^{\lambda}\left(q\right)$
for a gluon with helicity $\lambda=\pm$ is typically constructed 
using two null four vectors: a gluon momentum $k_{i}$ and a reference
momentum $q$ such that 
\begin{gather}
k_{i}\cdot\varepsilon_{i}^{\pm}\left(q\right)=q\cdot\varepsilon_{i}^{\pm}\left(q\right)=0,\label{eq:poldef_1a}\\
\varepsilon_{i}^{\pm}\left(q\right)\cdot\varepsilon_{i}^{\mp}\left(q'\right)=\varepsilon_{i}^{\pm}\left(q\right)\cdot\varepsilon_{i}^{\pm*}\left(q'\right)=-1.\label{eq:poldef_1b}
\end{gather}
For later use, let us note the following useful properties of the polarization
vectors
\begin{equation}
\varepsilon_{i}^{\pm}\left(q\right)\cdot\varepsilon_{i}^{\pm}\left(q'\right)=\varepsilon_{i}^{\pm}\left(q\right)\cdot\varepsilon_{j}^{\pm}\left(q\right)=\varepsilon_{i}^{\pm}\left(q\right)\cdot\varepsilon_{j}^{\mp}\left(k_{i}\right)=0,\label{eq:poldef_2a}
\end{equation}
where $k_{j}$ is a momentum of another gluon and $q'$ is some other
arbitrary null reference four vector. The change of a reference momentum
renders a vector proportional to the gluon momentum
\begin{equation}
\varepsilon_{i}^{\pm\mu}\left(q\right)=\varepsilon_{i}^{\pm\mu}\left(q'\right)+k_{i}^{\mu}\beta_{i}\left(q,q'\right),\label{eq:Refmom_change}
\end{equation}
where $\beta_{i}\left(q,q'\right)$ is a certain function which depends
on the actual representation of the polarization vectors (see also
 Eq.~(\ref{eq:LFpolvec3}) below). If an amplitude is gauge invariant,
the second term in (\ref{eq:Refmom_change}) does not contribute due
to the Ward identity.

One of the most convenient representations for the polarization vectors
is provided by the spinor formalism (see e.g. \cite{Mangano:1990by}).
However, for the purpose of this work we choose another representation,
given by 
\begin{equation}
\varepsilon_{i}^{\pm\mu}\left(\eta\right)\equiv\varepsilon_{i}^{\pm\mu}=\varepsilon_{\perp}^{\pm\mu}+\frac{\vec{\varepsilon}_{\perp}^{\,\,\pm}\cdot\vec{k}_{i\perp}}{k_{i}\cdot\eta}\eta^{\mu}\, ,\label{eq:LFpolvec1}
\end{equation}
with
\begin{equation}
\varepsilon_{\perp}^{\pm}=\frac{1}{\sqrt{2}}\left(0,1,\pm i,0\right).\label{eq:LFpolvec2}
\end{equation}
Here, the reference momentum is explicitly set to $q=\eta$ which
is especially convenient when one works in the light-like axial gauge
with the gauge vector $\eta$. The change of the reference momentum
can be realized as follows
\begin{equation}
\varepsilon_{i}^{\pm\mu}\left(q\right)=\varepsilon_{i}^{\pm\mu}\left(\eta\right)+\frac{1}{k_{i}\cdot q}\left(\vec{\varepsilon}_{\perp}^{\,\,\pm}\cdot\vec{q}_{\perp}-\frac{q\cdot\eta}{k_{i}\cdot\eta}\,\vec{\varepsilon}_{\perp}^{\,\,\pm}\cdot\vec{k}_{i\perp}\right)k_{i}^{\mu}.\label{eq:LFpolvec3}
\end{equation}

Throughout the paper we shall often encounter the following scalar
products
\begin{equation}
\tilde{v}_{ij}=\varepsilon_{j}^{-}\cdot k_{i}=k_{i}^{+}v_{ji} \, ,\label{eq:vijtild}
\end{equation}
where 
\begin{equation}
v_{ij}=\vec{\varepsilon}_{\perp}\cdot\left(\frac{\vec{k}_{i\perp}}{k_{i}^{+}}-\frac{\vec{k}_{j\perp}}{k_{j}^{+}}\right).\label{eq:vij}
\end{equation}
The quantities $\tilde{v}_{ij}$ and $v_{ij}$ satisfy several useful
relations which we list in Appendix \ref{sec:Identities}. In the following we shall frequently use the light-front longitudinal momentum fraction $z_{i}$ defined by
\begin{equation}
z_{i}\equiv\frac{k_{i}^{+}}{P^{+}}\,,
\end{equation}
with $P^{+}$ the total incoming longitudinal momentum. Since $P^+$ is the total momentum which is constant and all the objects are boost invariant we shall set $P^+=1$ for simplicity in the following. 
The variables $v_{ij}$ are related to the spinor products that are frequently used to express the helicity amplitudes in the literature
\begin{equation}
[ ji ] = \sqrt{2 z_i z_j} \, {\varepsilon}_{\perp}^{-} \cdot \left( \frac{\vec{k}_{i\perp}}{z_i}-\frac{\vec{k}_{j\perp}}{z_j}  \right) = \sqrt{2 z_i z_j} v_{ij} \, , \hspace*{0.3cm} \langle ij \rangle = \sqrt{2 z_i z_j} \, {\varepsilon}_{\perp}^{+} \cdot \left( \frac{\vec{k}_{i\perp}}{z_i}-\frac{\vec{k}_{j\perp}}{z_j}  \right) = \sqrt{2 z_i z_j} v_{ij}^* \, ,
\label{eq:spinorproduct}
\end{equation}
where the spinor products are defined as
\begin{equation}
\langle ij \rangle = \langle i-| j+ \rangle, \; \; \; \;  [ij] = \langle i+|j-\rangle ,
\end{equation}
and the chiral projections of the spinors for massless particles are defined as
\begin{equation}
|i\pm\rangle = \psi_{\pm}(k_i)=\frac{1}{2}(1\pm \gamma_5) \psi(k_i) \; ,  \;\;\; \langle \pm i | = \overline{\psi}_{\pm}(k_i) \; . 
\end{equation}
We shall use so called dual (or color) decomposition of amplitudes,
which is a standard technique to deal with multi-gluon amplitudes.
That is, for a gluonic amplitude with external colors $a_{1},\ldots,a_{N}$
the expansion has the form
\begin{equation}
\mathcal{M}^{a_{1}\ldots a_{2}}=\sum_{\left\{ 1,\ldots,N\right\} }\mathrm{Tr}\left(t^{a_{1}}\ldots t^{a_{N}}\right)\,\mathcal{M}\left(1,\dots,N\right),
\end{equation}
where the sum is over all non-cyclic permutations of the indices,
$t^{a}$ are the color generators and $\mathrm{Tr}$ denotes the trace
over colors. The so-called color-ordered amplitudes on the r.h.s contain only kinematical parts of the amplitude, and are built from 
only planar diagrams and their argument order indicate the order of
external legs. In what follows, we shall consider only one color-ordered
amplitude $\mathcal{M}\left(1,\dots,N\right)\equiv \mathcal{M}$.

\section{Recursion relation with off-shell light-front amplitudes}

\label{sec:RecRelation}
\begin{figure}
\centerline{\includegraphics[width=0.45\textwidth]{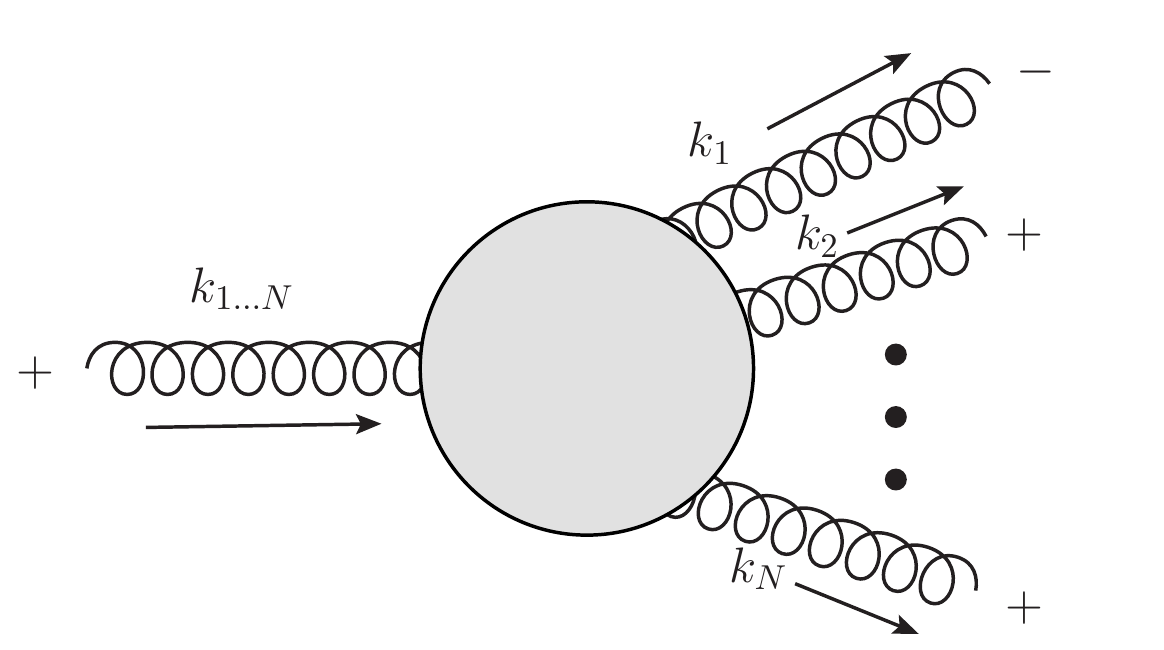}}
\caption{
\small
Diagram for the 1 to $N$ off-shell amplitude $\mathcal{M}_{N}$
for helicity configuration $\left(+\to-+\ldots+\right)$. The gluon with momentum $k_{1\dots N}$ is incoming and off-shell, whereas gluons with momenta $k_1, \dots, k_N$ are outgoing as indicated by the arrows.}
\label{fig:amp} 
\end{figure}
\begin{figure}
        \centering
        \begin{subfigure}[b]{0.5\textwidth}
                \includegraphics[width=\textwidth]{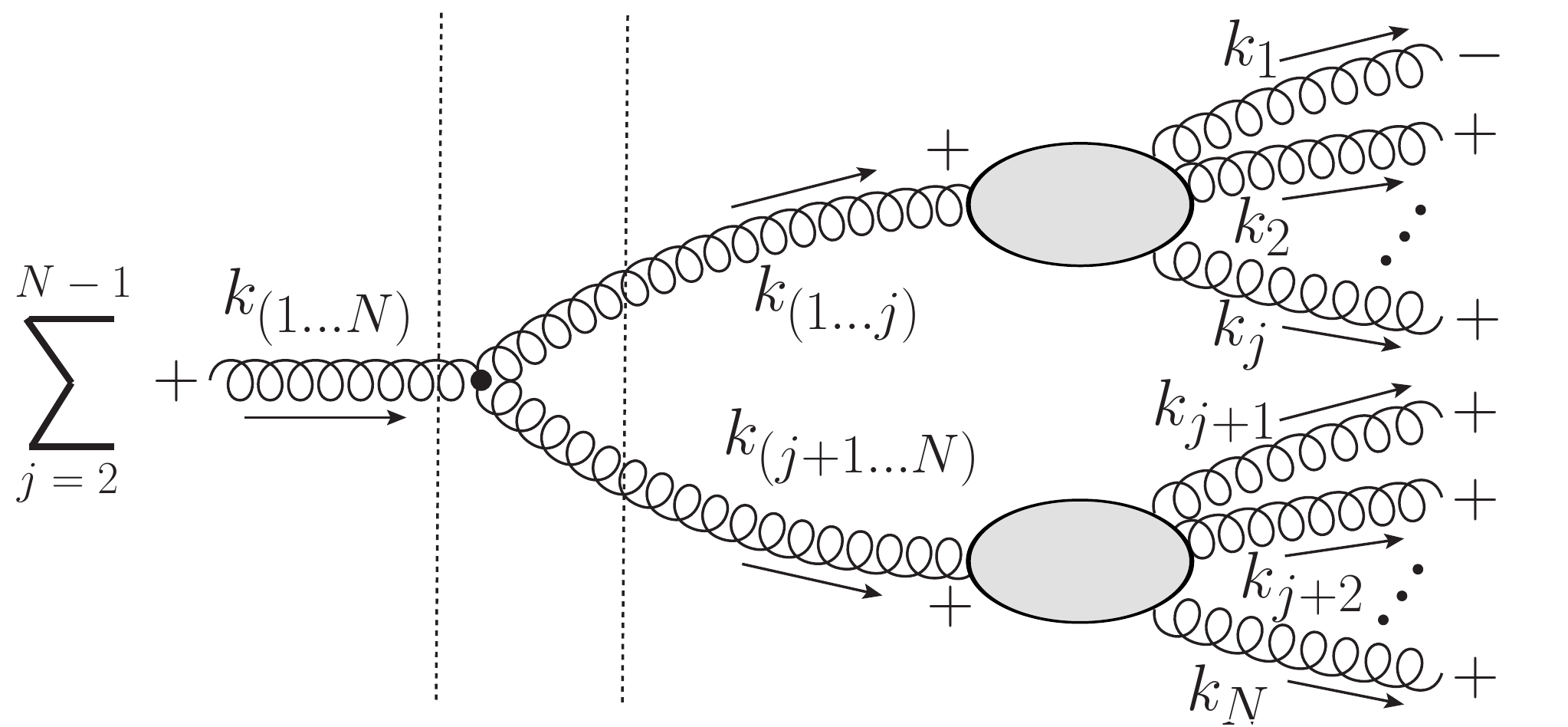}
                \caption{}
                \label{fig:gull}
        \end{subfigure}%
        ~ 
        \begin{subfigure}[b]{0.5\textwidth}
                \includegraphics[width=\textwidth]{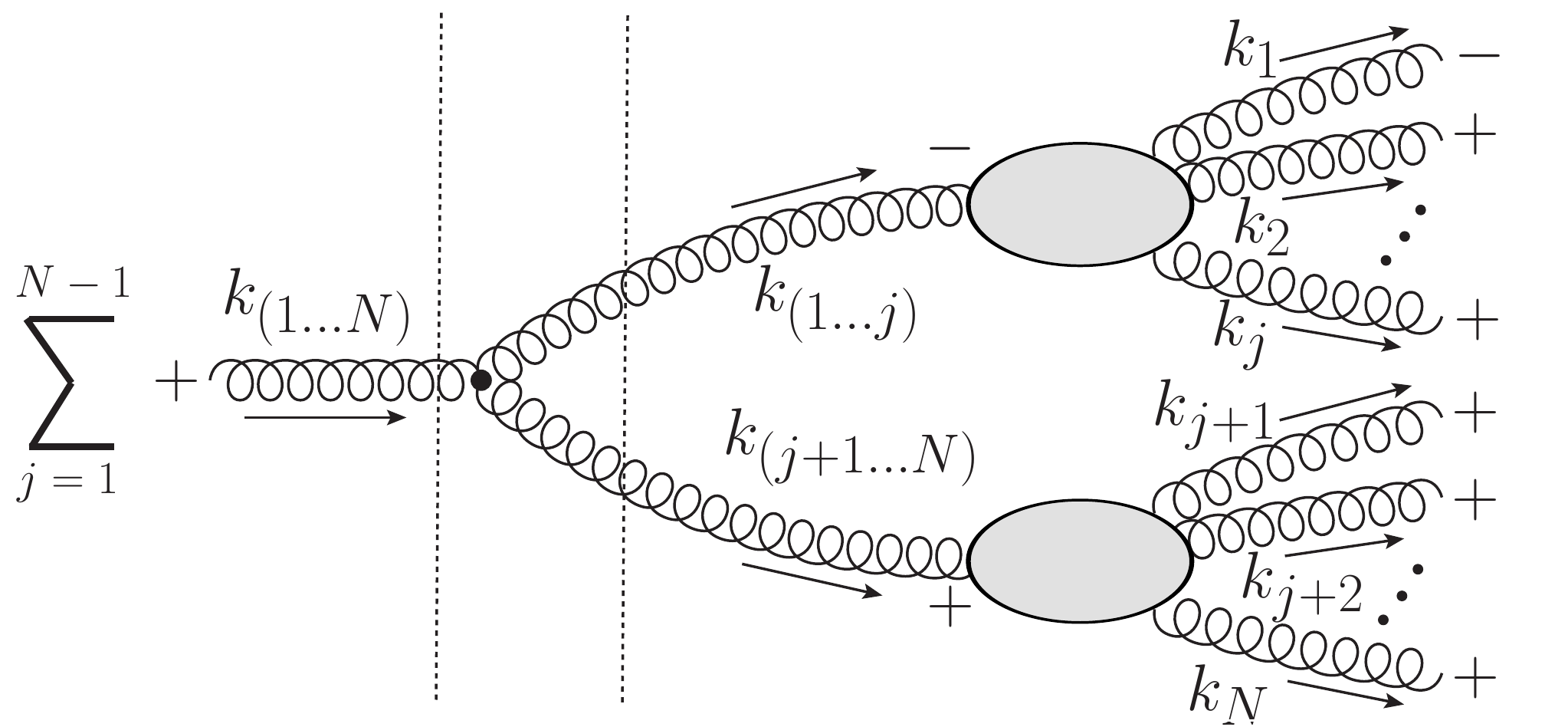}
                \caption{}
                \label{fig:tiger}
        \end{subfigure}
        ~ 
          
        \begin{subfigure}[b]{0.45\textwidth}
                \includegraphics[width=\textwidth]{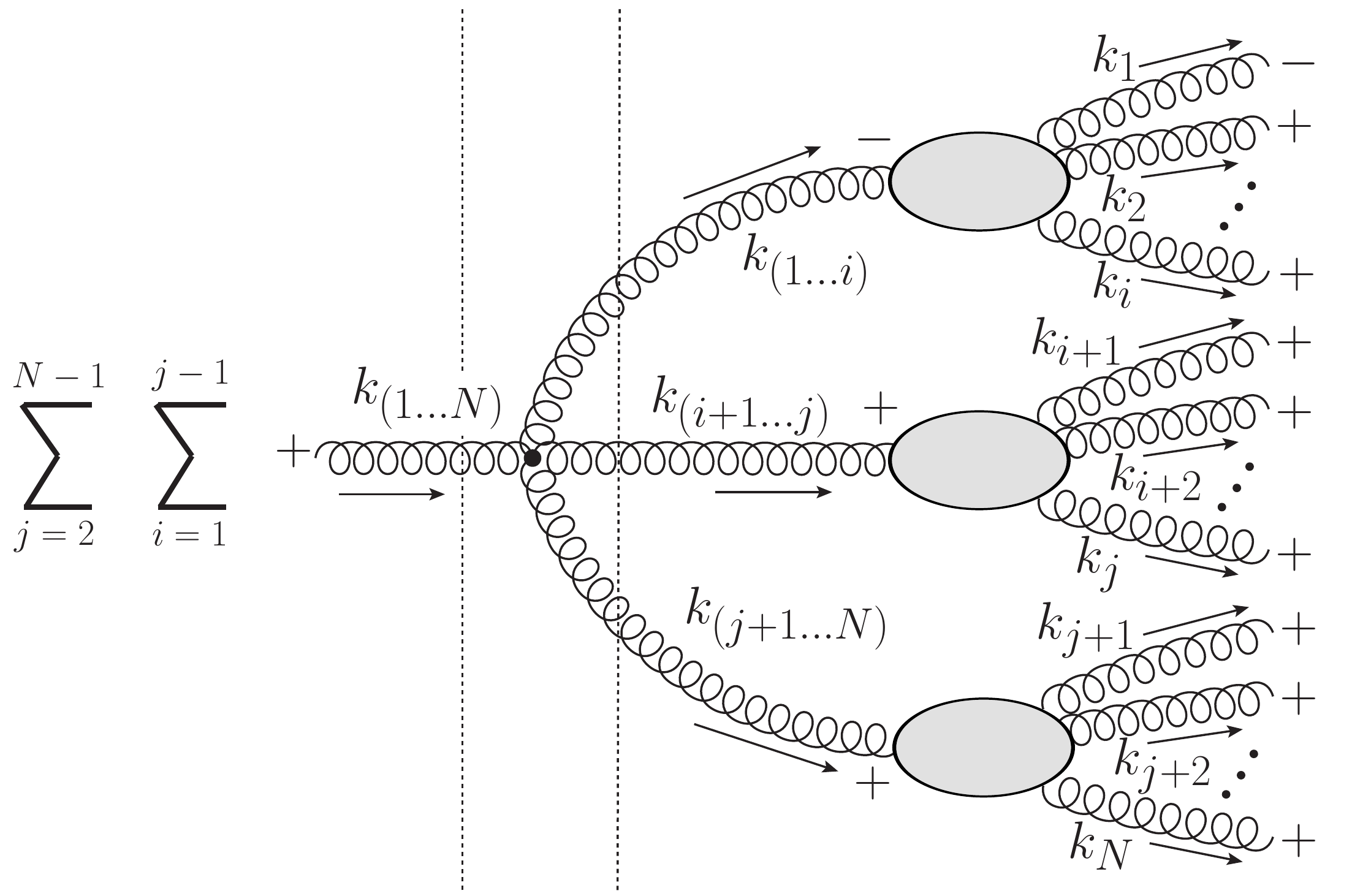}
                \caption{}
                \label{fig:mouse}
        \end{subfigure}
        ~ 
        \begin{subfigure}[b]{0.45\textwidth}
                \includegraphics[width=\textwidth]{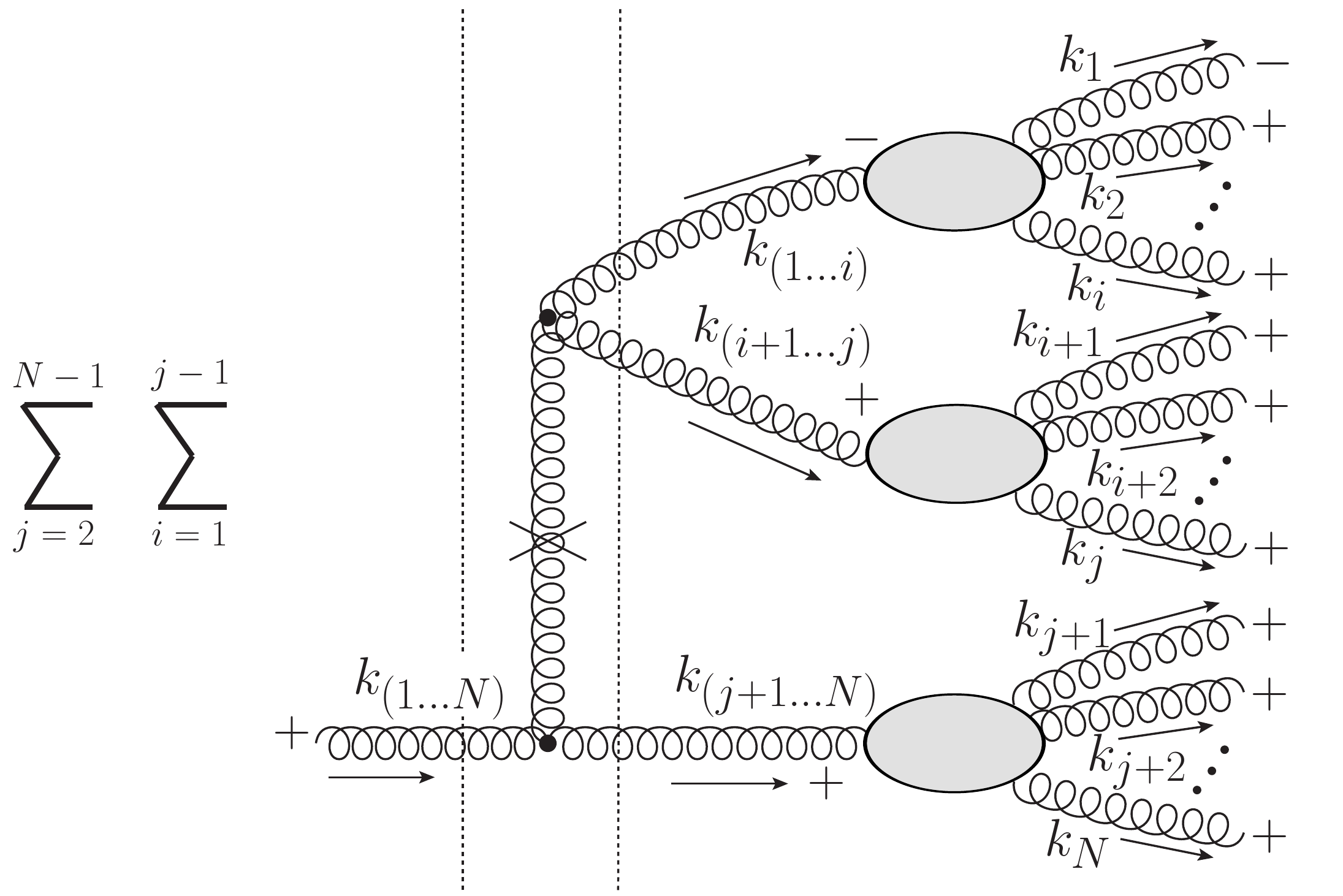}
                \caption{}
                \label{fig:mouse}
        \end{subfigure}

\caption{
\small
Schematic representation of the contributions to the off-shell amplitude $\mathcal{M}^{(+\rightarrow-+\dots+)}(k_{1\ldots N};k_{1},\ldots,k_{N})$. Graphs (a) and (b) involve contributions from the triple gluon vertex, graph (c) from the four-gluon vertex and graph (d) from the Coulomb interaction present in the light-front formalism. The sums on the left hand side of the graphs run over the number of the external legs in each of the contributing subamplitude in the graph. Graphs (a) and (b) differ by the helicity of the intermediate gluon which is incoming to the subamplitude $\mathcal{M}^{(\pm\rightarrow-+\dots+)}(k_{1\ldots j};k_{1},\ldots,k_{j})$. Vertical dotted lines indicate the energy denominators  which are present in the calculation. The summation over the light-front time orderings within the blobs has been already performed.}
\label{fig:recurgraphs}
\end{figure}

The color-ordered amplitude $\mathcal{M}$ is a QCD amplitude that is regularly obtained by the addition of the appropriate Feynman diagrams. The focus of this paper is on off-shell amplitudes which have a helicity configuration of $(+\to-+\ldots+)$, where the left-most particle is incoming and all the other particles are outgoing. In the on-shell limit of the incoming gluon this is by definition the
 Maximally Helicity Violating (MHV) amplitude which is the first non-trivial one\footnote{Typically for the helicity amplitudes, the convention used throughout the literature is that all the particles are outgoing. Therefore our amplitude $(+\rightarrow -+\dots+)$ in the on-shell limit corresponds to the MHV amplitude $(--+\dots+)$ which is indeed the first non-trivial one. The origin of the different convention in this paper stems from the fact that  we are performing calculations in the light-front perturbation theory with definite direction of the light-front time in the graphs. Also note that in the convention used in this paper '-' helicity corresponds to the '+'  (and vice versa) in the  convention used  in \cite{Mangano:1990by}.}, see for example \cite{Mangano:1990by}. By off-shell amplitudes we are referring to amplitudes which have incoming gluons off-shell and outgoing gluons on-shell and, henceforth, it should be assumed that all the processes and diagrams we discuss are off-shell unless otherwise specified. The particular process we will be looking at is depicted in Fig.~\ref{fig:amp}, where the left, incoming gluon with momentum $k_{1\ldots N}$ is the only off-shell gluon and where the single-line-blob
  represents the sum of all the possible intermediate processes. This off-shell amplitude, $\mathcal{M}^{(+\rightarrow-+\dots+)}(k_{1\ldots N};k_{1},\ldots,k_{N})$, has been previously studied using light-front  methods \cite{Cruz-Santiago:2013vta}.
To find this amplitude at the tree level for arbitrary number of external legs a recursion relation has been used
which is illustrated in Fig.~\ref{fig:recurgraphs}. Each of these graphs gives a contribution to this amplitude and it involves either triple vertex (graphs (a) and (b)), four-gluon vertex (graph (c)) or Coulomb interaction (graph (d)) as well as the amplitudes for lower number of legs and with different helicity configurations: $(+\to-+\ldots+), (+\to++\ldots+), (-\to-+\ldots+)$. Note that the last two amplitudes vanish in the on-shell limit, but, since the objects used here are off-shell, they are not zero, leading to non-trivial contributions to the recursion relation.  Finally, a summation over different combinations of the number of external partons is performed, as  illustrated in Fig.~\ref{fig:recurgraphs}. This recursion has been constructed using the factorization property of the fragmentation functions on the light-front, or the so-called cluster decomposition theorem \cite{Brodsky:1985gs}, and it is the light-front analog of the  
 Berends-Giele recursion relation \cite{Berends:1987me}.  We note that, to arrive at this recursion, the summation over all possible light-front time orderings of the vertices inside the blobs in Fig.~\ref{fig:recurgraphs} has been performed. 
 The solution to this recursion  relation yields the following expression for the off-shell amplitude 
\begin{multline}
\mathcal{M}^{(+\rightarrow-+\dots+)}(k_{1\ldots N};k_{1},\ldots,k_{N})= M^{(+\rightarrow-+\dots+)}(k_{1\ldots N};k_{1},\ldots,k_{N})-\\
z_{1\ldots N}^{2}D_{({1},\ldots,{N})}\sum_{i=2}^{N-1}\frac{1}{z_{1\ldots i}z_{1\ldots i+1}}\frac{1}{z_{i+1}\ldots z_{N}}\frac{g^{N-i}}{v_{i+1\; i+2}\ldots v_{N-1\; N}}\\
\frac{M^{(+\rightarrow-+\dots+)}(k_{1\ldots i};k_{1},\ldots,k_{i})}{v_{i+1(1\ldots i+1)}D_{({1},\ldots,{i})}}\;.\label{eq:masterrecursion}
\end{multline}
Here, $M$ is an object that appears when solving the recursion relation.  It has the same structure as the on-shell MHV amplitude but cannot be directly calculated from diagrams in the light-front. Given that its kinematics is
 the same as for $\mathcal{M}$, we refer to it as an off-shell amplitude as well. Later, in Sec.~\ref{sec:GI}, we will see that \eqref{eq:masterrecursion} can also be obtained from a
 matrix element of a straight infinite Wilson line. It will be then demonstrated  that, interestingly, $M$ is gauge invariant. Its explicit expression is\footnote{The normalization factor $\sqrt{2}^{j-1}$ in the MHV amplitude is due to the fact that in our convention the color ordered vertices do not contain $1/\sqrt{2}$ factors  and therefore they differ from convention used in \cite{Mangano:1990by}. }
 
\begin{align}
M^{(+\rightarrow-+\dots+)}(k_{1\ldots j};k_{1},\ldots,k_{j}) & \equiv -2ig^{j-1} \frac{z_{1\ldots j}\; z_{1}}{z_{2}z_{3}\ldots z_{j}}\frac{v_{(1\ldots j)1}^{3}}{v_{12}v_{23}\ldots v_{j-1\; j}v_{j(1\ldots j)}}  \nonumber \\
& = -i (\sqrt{2}g)^{j-1} \frac{[ (1\dots j)1 ]^4}{ [ (1\dots j) j] \, [ j (j-1) ]  \dots [  21 ] \, [ 1 (1\dots j)]} 
\;.\label{eq:Mdef}
\end{align}

 We should  note that, in order to have a fitting correspondence with \cite{Kotko:2014aba}, we use a different notation
for amplitudes 
  than \cite{Cruz-Santiago:2013vta}.  Furthermore, from now on we will be using the following shorthand notation for the arguments of the amplitudes: 
\begin{equation}
\mathcal{M} (k_{1\ldots j};k_{1\ldots i},k_{i+1},\ldots,k_{j})\equiv \mathcal{M} (k_{(1\ldots i)i+1\ldots j})\,.
\label{eq:shortnotation}
\end{equation}
This amplitude corresponds to a process $k_{1\dots j}\rightarrow k_{1\dots i} + k_{i+1}+\dots + k_{j}$.
For example, we denote  $\mathcal{M}^{(+\to+++)}(k_{123};k_{(12)},k_3)$ as $\mathcal{M}^{(+\to+++)}(k_{(12)3})$.

The energy denominator $D_{({1},\ldots,{i})}$ is defined as 
\begin{equation}
D_{({1},\ldots,{i})}=\sum_{j=1}^{i}E_{j}^{-}-E_{1\dots i}^{-}=\sum_{j=1}^{i}\frac{\vec{k}_{j\perp}^{2}}{z_{j}}-\frac{\vec{k}_{1\dots i\perp}^{2}}{z_{1\dots i}},
\end{equation}
it is the difference between the light-front energies of the outgoing
state and an intermediate state. Similarly energy denominator $D_{({1},\ldots,{N})}$
is 
\begin{equation}
D_{({1},\ldots,{N})}=\sum_{j=1}^{N}E_{j}^{-}-E_{1\dots N}^{-}=\sum_{j=1}^{N}\frac{\vec{k}_{j\perp}^{2}}{z_{j}}-\frac{\vec{k}_{1\dots N\perp}^{2}}{z_{1\dots N}}.
\end{equation}

One interesting aspect to
note about $\mathcal{M}^{(+\to-+\ldots+)}$ is that it is written as a sum of amplitudes $M^{(+\to-+\ldots+)}$ with different number of legs. Looking at Fig.~\ref{fig:recurgraphs}, it is not immediately obvious as to why $M^{(+\to-+\ldots+)}$ should dominate the expression since only Fig.~\ref{fig:recurgraphs}a has a substructure with helicity configuration $(+\to -+\ldots+)$. It turns out that the other graphs, even though their substructures do not have the right helicity configuration, do contribute terms proportional to $M^{(+\to-+\ldots+)}$.  Writing out the these terms   explicitly and through some algebraic manipulation, it can be shown \cite{Cruz-Santiago:2013vta} that  the following term emerges from Figs.~\ref{fig:recurgraphs}b-2d:
\[
\sum_{j=1}^{k}\frac{z_{1}z_{j+1\ldots k+1}^{2}}{z_{1\ldots j}}\frac{1}{v_{k+1\; k}\ldots v_{j+2\; j+1}v_{j\; j-1}\ldots v_{21}} \; \; .
\]
This term can be rewritten using the following  identity
\begin{multline}
-2i(-g)^{k}\frac{v_{k+1\;(1\ldots k+1)}}{z_{2}z_{3}\ldots z_{k}}\sum_{j=1}^{k}\frac{z_{1}z_{j+1\ldots k+1}^{2}}{z_{1\ldots j}}\frac{1}{v_{k+1\; k}\ldots v_{j+2\; j+1}v_{j\; j-1}\ldots v_{21}}\\
=z_{k+1}v_{k+1(1\ldots k+1)}\Bigg\{\frac{M^{(+\to-+\ldots+)}(k_{1\ldots k+1})}{v_{(1\ldots k+1)1}}\\
-\sum_{j=2}^{k}\frac{1}{z_{j+1}\ldots z_{k+1}}\frac{z_{1\ldots k+1}^{2}}{z_{1\ldots j}z_{1\ldots j+1}}\frac{1}{v_{j+1\; j+2}\ldots v_{k\; k+1}}\frac{g^{k+1-j}}{v_{j+1(1\ldots j+1)}}\frac{M^{(+\to-+\ldots+)}(k_{1\ldots j})}{v_{(1\ldots j)1}}\Bigg\}\;,\label{eq:relation}
\end{multline}
 and it is  seen that it is proportional to $M$.

The recursion relation \eqref{eq:masterrecursion} can be rewritten in a more elegant way, which demonstrates factorization into different subamplitudes. In order to do that  let us inspect the second term in \eqref{eq:masterrecursion}.
We shall show that it can be expressed as the sum over the products
$$\mathcal{M}^{(+\rightarrow+\dots+)}(k_{(1\ldots i) i+1\ldots N})\frac{i}{z_{1\dots i}D_{({1},\ldots,{i})}}M^{(+\rightarrow-+\dots+)}(k_{1\ldots i}) \; \; .$$
Let us start with the definition of the off-shell subamplitude for
the helicity configuration $(+\rightarrow+\dots+)$
\begin{equation}
\mathcal{M}^{(+\rightarrow+\dots+)}(k_{(1\ldots i) i+1\ldots N}) =-ig^{N-i}
 \frac{z_{1\dots N}^{2}}{z_{1\dots i}z_{i+1}\dots z_{N}}\frac{{D}_{({(1\dots i)},{i+1},\dots,{N})} }{v_{(1\ldots i)i+1}v_{i+1\; i+2}\ldots v_{N-1\; N}}\label{eq:mbarsub}.
\end{equation}
The energy denominator in this expression is equal to 
\begin{equation}
{D}_{({(1\dots i)},{i+1},\dots,{N})}=E_{1\dots i}^{-}+\sum_{j=i+1}^{N}E_{j}^{-}-E_{1\dots N}^{-}=\frac{\vec{k}_{1\dots i\perp}^{2}}{z_{1\dots i}}+\sum_{j=i+1}^{i}\frac{\vec{k}_{j\perp}^{2}}{z_{j}}-\frac{\vec{k}_{1\dots N\perp}^{2}}{z_{1\dots N}}.
\end{equation}
Note that \eqref{eq:mbarsub} vanishes in the on-shell limit
${D}_{({(1\dots i)},{i+1},\dots,{N})}=0$, which is consistent
with the fact that the on-shell amplitude vanishes for this helicity
configuration $(+\rightarrow+\dots+)$. Comparing with \eqref{eq:masterrecursion},
we see that the term inside the sum is very similar to the above off-shell
subamplitude with some additional prefactors.

Substituting \eqref{eq:mbarsub} in \eqref{eq:masterrecursion}
and using the identity (\ref{eq:Ident5a}) 
we finally obtain the following version of the recursion relation
\begin{multline}
\mathcal{M}^{(+\rightarrow-+\dots+)}(k_{1\ldots N})
=M^{(+\rightarrow-+\dots+)}(k_{1\ldots N}) \\
+\sum_{i=2}^{N-1}\frac{D_{({1},\ldots,{N})}}{{D}_{({(1\dots i)},{i+1},\dots,{N})}}\mathcal{M}^{(+\rightarrow+\dots+)}(k_{(1\ldots i) i+1\ldots N})\frac{i}{z_{1\dots i}D_{({1},\ldots,{i})}}M^{(+\rightarrow-+\dots+)}(k_{1\ldots i})\;.\label{eq:masterrecursion2}
\end{multline}

\begin{figure}
\centerline{\includegraphics[width=12cm]{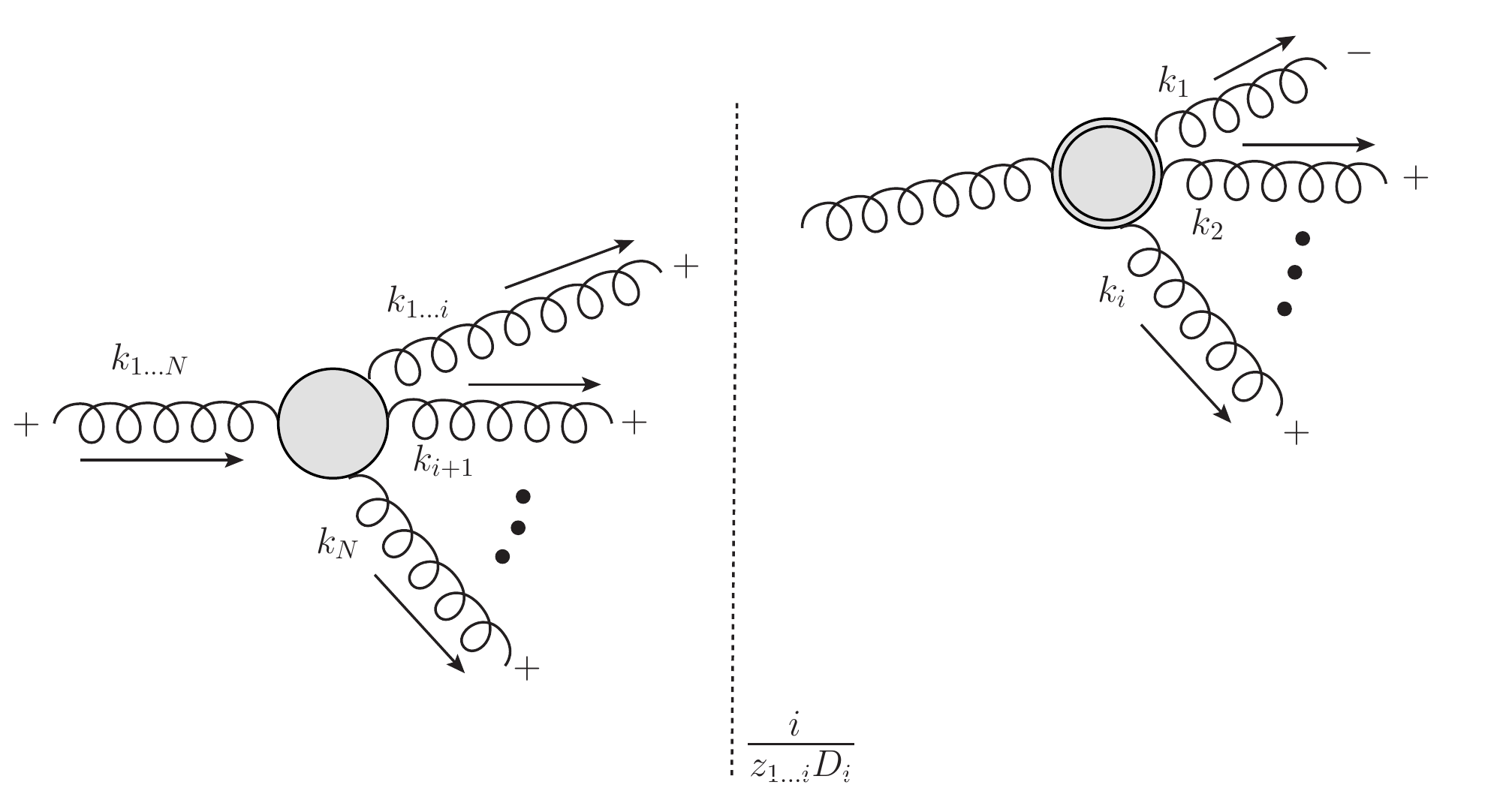}}
\caption{
\small
Schematic representation of the second term in the recursion formula
\eqref{eq:masterrecursion2}. The dotted vertical line  represents the energy denominator 
$1/{D_i}$,
 the graph on the left hand side of this line is the amplitude $\mathcal{M}^{(+\rightarrow+\dots+)}(k_{(1\ldots i) i+1\ldots N})$ whereas the graph on the right hand side is the amplitude $M^{(+\rightarrow-+\dots+)}(k_{1\ldots i})$. The double line around the blob in  the latter graph indicates that this is $M$ (with an explicit form of \eqref{eq:Mdef}) rather than $\mathcal{M}$ which are different objects as  explained in the text.}
\label{fig:fact2}
\end{figure}

We see that the second term  on the right hand side of this recursion has a nice factorized form
which can be recast diagrammatically as in Fig.~\ref{fig:fact2}. It consists of
the sum over the factorized products of amplitudes $M$
and~$\mathcal{M}$. $\mathcal{M}$,~however, is evaluated
with a different denominator in the sense that the ratio 
$D_{({1},\ldots,{N})}/{D}_{({(1\dots i)},{i+1},\dots,{N})}$
 cancels the energy denominator inside of $\mathcal{M}$ and replaces it with $D_{({1},\ldots,{N})}$.
We shall prove in Sec.~\ref{sec:GI} that the above recursion relation has its roots in the 
recurrence property of the straight infinite Wilson line, which  involves gauge invariant amplitude  $M$.

\section{The Ward identity for light-front amplitudes}

\label{sec:WardId}

Since, in the present work, we are interested in the gauge invariance
properties of the off-shell amplitudes on the light front, in this
section we want to discuss certain issues regarding the Ward identities within this framework. 
In order to illustrate the issue, we shall first 
 verify the identity for the lowest order amplitude $(+\rightarrow -++)$ on the light-front.
It will become clear that one needs to modify the rules for the 
computation of the Ward identities, 
in order to 
guarantee the four-momentum conservation. Standard light-front rules do not involve the minus components
in the calculation, except for the denominators, where it is not conserved \cite{Kogut:1969xa,Lepage:1980fj,Brodsky:1997de}. However, for the Ward identity to hold, 
the minus components need to be taken care of in the vertices as well and, thus, the procedure for the computation
of this identity needs to be revised.
We shall demonstrate that this results in the additional instantaneous-like component, which needs to be taken into account. Then   the result of the calculation is proportional to the energy denominator  of the initial  state and  the Ward identity holds on the light-front.
Second, we shall perform the Ward identity check using  the recursion relation \eqref{eq:masterrecursion}. It turns out that the second term in the r.h.s. of \eqref{eq:masterrecursion2}, which is a sum of lower order amplitudes in this recursion, gives the expression which is exactly equal to the term previously derived by the explicit calculation of the Ward identity from diagrams. This means that the new 
amplitude $M$ which appears in the recursion relation is gauge invariant, i.e. the Ward identity gives exactly zero for this object despite the fact that it is off-shell.

\subsection{Example: the Ward identity check for the lowest order amplitude}

Let us recall, that for a generic QCD amplitude $\mathcal{M}$ with
external momenta $k_{i}$ on-shell and corresponding polarization
vectors $\varepsilon_{i}$, the Ward identities read
\begin{equation}
\left.\mathcal{M}\right|_{\varepsilon_{i}\rightarrow k_{i}}=0\,\,\,\,\,\textrm{for any }i\, .\label{eq:WardIdentity}
\end{equation}
These identities do not work for light-front amplitudes when applied directly and appropriate modifications must be performed to ensure that they are satisfied.
The reason is that one injects into the vertices a minus light cone component
when replacing a gluon polarization vector by its momentum, while
the actual minus light cone components flowing through the diagram
are integrated out \textit{prior} to this replacement. 

\begin{figure}
\begin{centering}
\includegraphics[width=0.5\paperwidth]{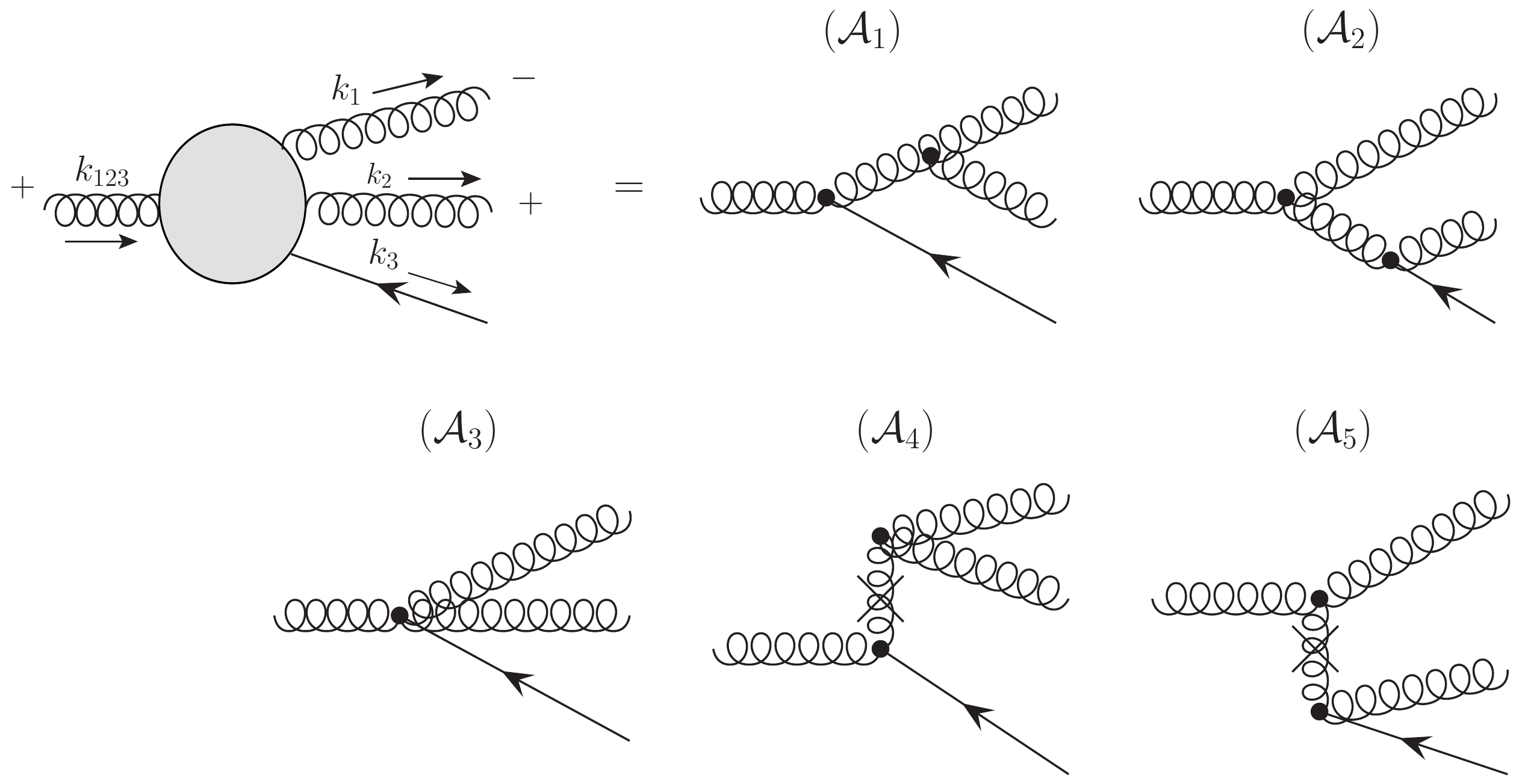}
\par\end{centering}
\caption{\small Diagrams for the Ward identity check for $1\rightarrow3$
light-front amplitude with helicity configuration $+\rightarrow-+\pm$.
The solid line with an arrow instead of a gluon line represents replacement
of a polarization vector with the corresponding momentum. The two
right-most diagrams contain instantaneous interactions.\label{fig:M1to3WI}}

\end{figure}

The problem can be illustrated by the following explicit example. Consider a
light-front amplitude for the $1\rightarrow3$ process with the helicity configuration
$\left(+\rightarrow-++\right)$ (first particle is incoming and the rest 
are outgoing). This is the lowest non-trivial MHV amplitude. We then replace the  polarization
vector of the third outgoing particle by the corresponding momentum. We have
\begin{equation}
\mathcal{M}_{1\rightarrow3}^{\left(+\rightarrow-+k_{3}\right)}=\mathcal{A}_{1}+\mathcal{A}_{2}+\mathcal{A}_{3}+\mathcal{A}_{4}+\mathcal{A}_{5},\label{eq:M1to3WI1}
\end{equation}
where $\mathcal{A}_{1}$-$\mathcal{A}_{5}$ are the contributions
from the diagrams depicted in Fig.~\ref{fig:M1to3WI}. 
Here, and below we use a notation for replacement $\epsilon_i\leftrightarrow k_i$ in the superscript, i.e. we replace the helicity indication by the corresponding momentum.
Using the rules of the LFPT and color-ordered vertices  we get
\begin{align}
\mathcal{A}_{1}& =-2ig^{2}\,\frac{z_{2}z_{3}z_{123}\left(z_{123}+z_{12}\right)}{z_{12}^{2}D_{(1,2)}}\, v_{\left(12\right)1}v_{\left(123\right)3}v_{\left(123\right)3}^{*} \, ,\label{eq:D1} \\
\mathcal{A}_{2} & =-2ig^{2}\,\frac{z_{2}z_{3}z_{23}\left(z_{2}+z_{23}\right)}{z_{23}^{2}D_{(2,3)}}\, v_{\left(123\right)1}v_{23}v_{23}^{*}\, ,\label{eq:D2} \\ 
\mathcal{A}_{3} & =ig^{2}\, z_{3}\left(v_{\left(123\right)3}-2v_{13}\right)\, ,\label{eq:D3} \\
\mathcal{A}_{4} & =ig^{2}\,\frac{z_{3}\left(z_{1}-z_{2}\right)}{z_{12}}\, v_{\left(123\right)3}\, ,\label{eq:D4} \\
\mathcal{A}_{5}& =0\, .\label{eq:D5}
\end{align}
Let us note that the above results are obtained without the full
four-momentum conservation as, according to LFPT rules, at each vertex
there is a Dirac delta for the plus and transverse components,
but not for minus components, which for each momentum are fixed
by the on-shell condition. Adding the diagrams we get
\begin{multline}
\mathcal{M}_{1\rightarrow3}^{\left(+\rightarrow-+k_{3}\right)}=-ig^{2}\bigg[2\frac{z_{3}}{z_{12}}\left(z_{1}v_{1\left(123\right)}+z_{2}v_{13}\right)+\left(z_{23}+z_{2}\right)v_{\left(123\right)1}\\
-\frac{z_{2}z_{3}z_{123}}{z_{1}z_{12}^{2}v_{12}^{*}}\left(z_{123}+z_{12}\right)v_{\left(123\right)3}v_{\left(123\right)3}^{*}\bigg].\label{eq:M1to3WI2}
\end{multline}
If the Ward identity was satisfied, this result should be proportional
to the energy denominator $D_{(1,2,3)}$, which vanishes for
physical on-shell partons. However this is not the case for (\ref{eq:M1to3WI2}). 

The above problem stems from the fact that the light-front diagrams in Fig.~\ref{fig:M1to3WI} were obtained assuming that there are no external
minus components. Indeed, for an amplitude calculation, the polarization
vectors (\ref{eq:LFpolvec1}) project only on plus and transverse
components. The only minus components that flew inside diagrams
were integrated out giving energy denominators and instantaneous terms.
However, for the above Ward identity check, the triple gluon vertex
that appears, for example, in $\mathcal{A}_{1}$, reads
\begin{multline}
V_{3}^{+\rightarrow+k_{3}}\left(-k_{23},k_{2},k_{3}\right)=ig\Bigg[-k_{3}\cdot\left(k_{2}+k_{23}\right)\left(\varepsilon_{23}^{+}\right)^{*}\cdot\varepsilon_{2}^{+}+\left(\varepsilon_{23}^{+}\right)^{*}\cdot\left(k_{2}-k_{3}\right)k_{3}\cdot\varepsilon_{2}^{+}\\
+\varepsilon_{2}^{+}\cdot\left(k_{23}+k_{3}\right)\left(\varepsilon_{23}^{+}\right)^{*}\cdot k_{3}\Bigg]=ig\,\frac{z_{2}z_{3}\left(z_{23}+z_{2}\right)}{z_{23}}v_{23}v_{23}^{*}.\label{eq:V3a}
\end{multline}
The problematic term is the first one in the square bracket. Formally,
we cannot write $k_{3}\cdot\left(k_{2}+k_{23}\right)=2k_{2}\cdot k_{3}$
since we do not have full momentum conservation. We have to consider
$k_{2}\cdot k_{3}$ and $k_{23}\cdot k_{3}$ as different scalar products
and this causes the Ward identity to fail.

As already mentioned in the beginning of this section, in the correct
procedure one should integrate out all the minus components.
In (\ref{eq:V3a}) the internal minus component $k_{23}^{-}$
appears in the numerator and thus leads to instantaneous-like additional
contribution to (\ref{eq:M1to3WI1}). We give an explicit calculation
of this term in Appendix \ref{sec:App_WI}. It turns out, that this
additional term added to (\ref{eq:V3a}) gives 
\begin{equation}
\tilde{V}_{3}^{+\rightarrow+k_{3}}\left(-k_{23},k_{2},k_{3}\right)=2ig\, z_{2}z_{3}v_{23}v_{23}^{*} \; ,\label{eq:V3b}
\end{equation}
and it effectively restores the full momentum conservation (in the numerator). Repeating the procedure for $\mathcal{A}_{2}$ (the
other diagrams are not affected) we get
\begin{equation}
\mathcal{M}_{1\rightarrow3}^{\left(+\rightarrow-+k_{3}\right)}=ig^{2}\,\frac{z_{2}}{z_{1}}\,\frac{D_{\left(1,2,3\right)}}{v_{12}^{*}} \; ,\label{eq:M1to3WI3}
\end{equation}
which obviously vanishes in the case of the on-shell amplitude, i.e. for $D_{\left(1,2,3\right)}\rightarrow0$.

\subsection{Ward identity and the recursion relation for the lowest order amplitude}

In the recursion relation \eqref{eq:masterrecursion} the new amplitude  $M$ that actually solves the recurrence, has exactly 
the form of the MHV amplitude. Therefore, once we impose the on-shell condition for the process, the result is equal to 
the MHV amplitude as expected. The amplitude $M$, however, as it stands in the recursion relation, is an off-shell object.
As we shall see shortly, it has a remarkable property, namely, it turns out that it is gauge invariant.

This can be explicitly illustrated by taking (\ref{eq:masterrecursion2})
for $n=3$ and checking the Ward identity.  However, one needs to take care of the issues
discussed in the preceding section. We need to calculate
\begin{equation}
M_{1\rightarrow3}^{\left(+\rightarrow-+k_{3}\right)}=\mathcal{M}_{1\rightarrow3}^{\left(+\rightarrow-+k_{3}\right)}-\frac{D_{\left(1,2,3\right)}}{D_{\left(12,3\right)}}\mathcal{M}_{1\rightarrow2}^{\left(+\rightarrow+ k_3\right)}\frac{i}{z_{12}D_{\left(1,2\right)}}\, M_{1\rightarrow2}^{\left(+\rightarrow-+\right)}.\label{eq:GI_rec1to3}
\end{equation}
In the above expression we have replaced the polarization vector both in the $1 \to 3$ amplitude and $1 \to 2$ subamplitude.
The second term can be simplified to
\begin{equation}
\frac{D_{\left(1,2,3\right)}}{D_{\left(12,3\right)}}\left(2ig\, z_{3}z_{123}v_{\left(123\right)3}v_{\left(123\right)3}^{*}\right)\frac{i}{z_{12}D_{\left(1,2\right)}}\left(2igz_{2}v_{\left(12\right)1}\right)=ig^{2}\,\frac{z_{2}}{z_{1}}\frac{D_{\left(1,2,3\right)}}{v_{12}^{*}}\label{eq:GI_2}
\end{equation}
where we have used the relations $z_{12}D_{\left(1,2\right)}=2z_{1}z_{2}v_{12}v_{12}^{*}$,
$z_{12}D_{\left(12,3\right)}=2z_{123}z_{3}v_{\left(123\right)3}v_{\left(123\right)3}^{*}$
and $z_{1}v_{\left(12\right)1}=-z_{2}v_{12}$. We see that (\ref{eq:GI_2})
precisely cancels the previously derived term (\ref{eq:M1to3WI3}), leaving $M_{1\rightarrow3}^{\left(+\rightarrow-+k_{3}\right)}$ equal to zero.
Therefore $M_{1\rightarrow3}^{\left(+\rightarrow-++\right)}$ is the gauge invariant amplitude irrespectively whether  the incoming leg is on-shell or off-shell.

It may be argued that the Ward identity for $M$ is satisfied in general, for arbitrary number of external legs
\begin{equation}
M_{1\rightarrow N}^{\left(+\rightarrow-+\dots k_{i}\dots+\right)}=0.\label{eq:M1tonWI}
\end{equation}
We shall undertake this task in the next section.

\section{Proof of gauge invariance of the amplitude $M$ from Wilson lines}

\label{sec:GI}

In the previous section it has been claimed that the off-shell amplitude $M$ which appears in the recurrence relation (\ref{eq:masterrecursion})
is gauge invariant and thus satisfies the Ward identities (\ref{eq:M1tonWI}). 
Although, in principle, one could show this by arranging a similar recurrence for the Ward identity and showing that $M$ vanishes for arbitrary number of legs, we will study a connection of equation (\ref{eq:masterrecursion}) with the matrix element of certain straight infinite Wilson line (or gauge link) operator. For that object one can immediately write a recurrence which resembles (\ref{eq:masterrecursion}), which,  after a careful derivation turns out to be exactly  the same. We will start by reviewing the Wilson line approach for off-shell amplitudes. Later, we will derive the recursion (\ref{eq:masterrecursion}) directly from this approach.

\subsection{Matrix elements with Wilson lines and off-shell amplitudes}

\label{subsec:Wilsonlines}

Let us consider a tree level gluonic Green's function in momentum space with external momenta $k_{1\dots N},k_1,\dots,k_N$ satisfying momentum conservation (we assume, as before, that $k_{1\dots N}$ is incoming and the rest are outgoing). As such, the Green's function is a purely off-shell object, i.e. the external momenta have arbitrary virtuality; moreover, the external gluon Lorentz indices are not contracted. In order to obtain a scattering amplitude, we reduce the Green's function by amputating the external propagators, taking the on-shell limit for the external momenta, and contracting the external legs with appropriate polarization vectors transverse to (on-shell) momenta. Here, we shall consider the Green's function where the legs $k_1,\dots,k_N$ are on-shell and reduced as above, while the leg $k_{1\dots N}$ is kept off-shell and is contracted with a vector $\mathfrak{e}_{1\dots N}$. We shall call this vector a ``polarization'' vector for the off-shell gluon. At this point, it is only assumed that this vector is transverse to the off-shell momentum, $\mathfrak{e}_{1\dots N}\cdot k_{1\dots N}=0$. We will call the Green's function reduced in that manner an off-shell amplitude.

The off-shell amplitude constructed according to the above procedure is not gauge invariant, i.e. it does not satisfy the Ward identities with respect to the on-shell legs (for a general choice of $\mathfrak{e}_{1,\dots,N}$ and external polarization vectors). However, one can find a gauge invariant extension of such  off-shell amplitude.  For example, in analysis of scattering at high-energy  one encounters similar objects. There, the $\mathfrak{e}_{1\dots N}$ is set to 
 one of the 
 light-cone components $n_{\pm}$ ($n^2_{\pm}=0$) of a hadron momentum and $k_{1\dots N}=xn_{\pm}+k_T$, so that $k_{1\dots N}\cdot n_{\pm}=0$. The gauge invariant vertices corresponding to transitions of such off-shell gluons to a set of on-shell gluons can be derived from the so-called Lipatov's effective action \cite{Lipatov:1995pn,Antonov:2004hh}.

In Ref.~\cite{Kotko:2014aba} the author discussed a more general situation, where $\mathfrak{e}_{1\dots N}$ is arbitrary. In that case the gauge invariant extension of the off-shell amplitude can be found by 
considering a matrix element of a straight infinite Wilson line operator. More precisely, one defines an object
\begin{multline}
\mathfrak{M}_{\mathfrak{e}_{1\dots N}}^{a_{1\dots N}a_{1}\ldots a_{N}}\left(k_{1\dots N};k_{1},\ldots,k_{N}\right)=\int\!\! d^{4}x\, e^{ik_{1\dots N}\cdot x}\\
\left\langle 0\left|\mathcal{T}\left\{ R_{\mathfrak{e}_{1\dots N}}^{\,a_{1\dots N}}(x)\, e^{iS_{\,\textrm{Y-M}}}\right\} \right|k_{1},\lambda_{1},a_{1};\ldots;k_{N},\lambda_{N},a_{N}\right\rangle _{c}\, ,\label{eq:Mfrak_1-1}
\end{multline}
with
\begin{equation}
R_{\mathfrak{e}_{1\dots N}}^{\,a_{1\dots N}}(x) = \,\mathrm{Tr}\left[t^{a_{1\dots N}}\mathcal{P}\exp\left(ig\int_{-\infty}^{+\infty}ds\, A_{\mu}^{b}\left(x+s\,\mathfrak{e}_{1\dots N}\right)\mathfrak{e}_{1\dots N}^{\mu}t^{b}\right)\right]\, ,
\label{eq:Roperator}
\end{equation}
where $\mathcal{T}$ is the time-ordering, $\mathcal{P}$ is the path-ordering,
$S_{\,\textrm{Y-M}}$ is the Yang-Mills interaction action, and, finally,
$\left|k_{i},\lambda_{i},a_{i}\right\rangle $ are one-gluon on-shell
states with momentum $k_{i}$, helicity $\lambda_{i}$ and color $a_{i}$.
The color of the off-shell gluon is $a_{1\dots N}$. The subscript
$c$ means that we take only connected contributions. The infinite
Wilson line operator $R_{\mathfrak{e}_{1\dots N}}^{\,a_{1\dots N}}$  sandwiched in the matrix element is explicitly
gauge invariant with respect to small gauge transformations. Actually, in Ref.~\cite{Kotko:2014aba}, instead of a straight infinite path in (\ref{eq:Roperator}), deformed paths were considered in order to regularize the integrals and to show that they form certain  generalized functions.

Let us now consider a color-ordered version of the matrix element (\ref{eq:Mfrak_1-1})
with order $\left(a_{1\dots N},a_{1},\dots,a_{N}\right)$.
According to  \cite{Kotko:2014aba} it is proportional to the momentum conservation Dirac delta and the
delta assuring the Wilson line direction $\mathfrak{e}_{1\dots N}$
and the momentum $k_{1\dots N}$ are mutually transverse
\begin{equation}
\mathfrak{M}_{\mathfrak{e}_{1\dots N}}\left(k_{1\dots N}\right)=\delta^{4}\left(k_{1\dots N}-k_{1}-\ldots-k_{N}\right)\delta\left(\mathfrak{e}_{1\dots N}\cdot k_{1\dots N}\right)\tilde{\mathcal{M}}_{\mathfrak{e}_{1\dots N}}^{\left(\lambda_{1}\dots\lambda_{N}\right)}\left(k_{1\dots N}\right)\, ,\label{eq:Mfrak_2}
\end{equation}
where we have used the shorthand notation for momenta arguments as defined in Eq.~(\ref{eq:shortnotation}).
The above relation defines the \textit{gauge invariant off-shell amplitude}
$\tilde{\mathcal{M}}$ with ``polarization'' vector $\mathfrak{e}_{1\dots N}$
for the off-shell gluon. It satisfies the Ward identities with respect
to the external on-shell legs (but not with respect to $\mathfrak{e}_{1\dots N}$, i.e. 
the Wilson line slope)
\begin{equation}
\tilde{\mathcal{M}}_{\mathfrak{e}_{1\dots N}}^{\left(\lambda_{1}\dots k_{i}\dots\lambda_{N}\right)}\left(k_{1\dots N}\right)=0\,\,\,\,\mathrm{for}\,\,\, i=1,\ldots,N\, .\label{eq:Mtild_1}
\end{equation}
 Let us stress that the amplitude $\tilde{\mathcal{M}}$ is indeed gauge invariant only when $k_{1\dots N}\cdot \mathfrak{e}_{1\dots N}=0$.

Diagrammatically, the amplitude $\tilde{\mathcal{M}}$
can be written as

\begin{tabular}{>{\centering}m{0.83\columnwidth}>{\centering}m{0.05\columnwidth}}
\bigskip{}
\centering{}\includegraphics[height=0.16\paperheight]{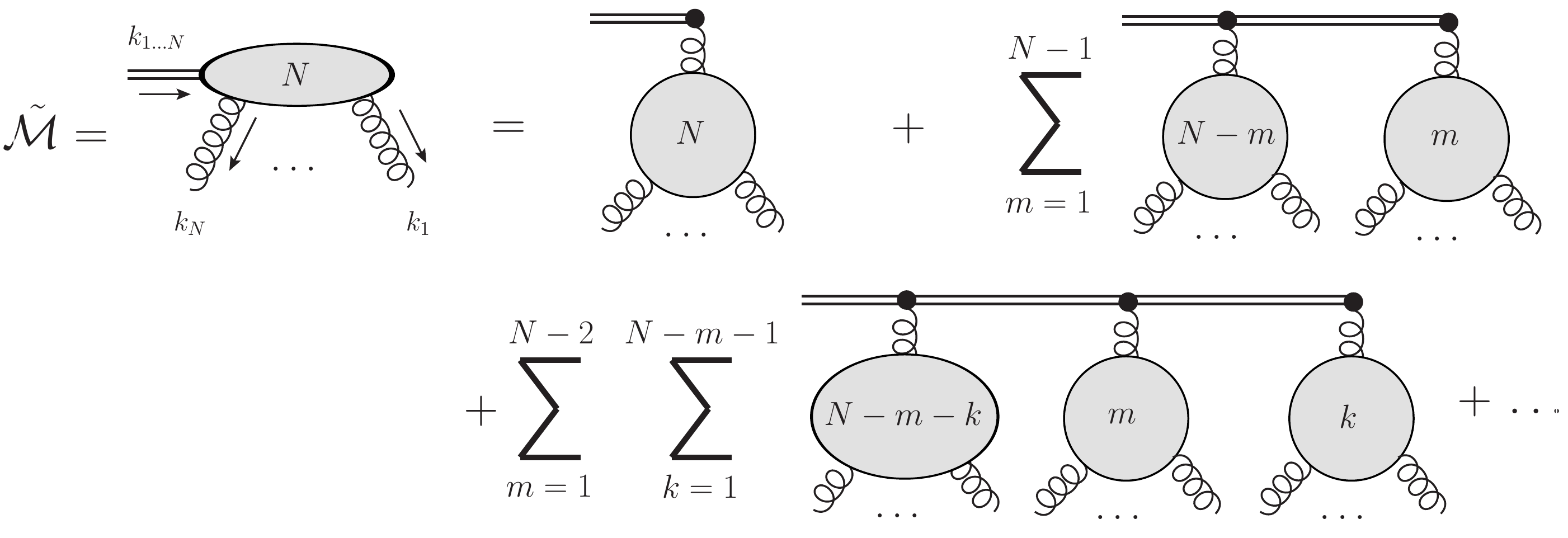} & \centering{}(\myref)\label{eq:Gaugelink_expansion}\tabularnewline
\end{tabular}

\noindent The double line represents the Wilson line in momentum space. Each double
line  connecting two gluon attachments contributes the propagator
$i/p\!\cdot\!\mathfrak{e}_{1\dots N}$, with $p$ being the momentum flowing
through the line. The gluons couple to the Wilson line via an $ig\mathfrak{e}_{1\dots N}^{\mu}$
vertex. More on the Feynman rules can be found in \cite{Kotko:2014aba}
\footnote{We have modified the rules of \cite{Kotko:2014aba} by ``cutting
off'' the double line carrying the zero momentum, which is more transparent.%
}. The blobs represent standard QCD contributions with the numbers indicating
the number of external on-shell legs. The ellipses after the last plus sign represent
contributions with more blobs connected to the gauge link. 
Note that
the first contribution in (\ref{eq:Gaugelink_expansion}) is the off-shell amplitude defined at the beginning of this subsection (modulo $ig$ factor due to a coupling with the gauge link). In what follows we will denote this amplitude as $\overline{\mathcal{M}}^{\left(\mathfrak{e}_{1\dots N}\rightarrow\lambda_{1}\dots\lambda_{N}\right)}\left(k_{1\dots N}\right)$. It will contain the off-shell propagator and and a coupling to the Wilson line 
 (we include an additional $i$ factor for further convenience)
\begin{equation}
\overline{\mathcal{M}}^{\left(\mathfrak{e}_{1\dots N}\rightarrow\lambda_{1}\dots\lambda_{N}\right)}\left(k_{1,\dots,N}\right)=ig\frac{-i}{k_{1\dots N}^{2}}\,\,\,i\mathcal{M}^{\left(\mathfrak{e}_{1\dots N}\rightarrow\lambda_{1}\dots\lambda_{N}\right)}\left(k_{1,\dots,N}\right)\, ,\label{eq:MbarQCDdef}
\end{equation}
where $\mathcal{M}$ is the standard QCD amplitude calculated from Feynman diagrams (with, however, off-shell kinematics). 
Let us underline one more time that the amplitude $\mathcal{M}$ (or $\overline{\mathcal{M}}$) itself does not satisfy the Ward identities,
but they are restored thanks to the rest of the r.h.s of Eq.~(\ref{eq:Gaugelink_expansion}).

The decomposition (\ref{eq:Gaugelink_expansion}) can be written in
a more compact form by means of the following recursion:

\begin{tabular}{>{\centering}m{0.83\columnwidth}>{\centering}m{0.05\columnwidth}}
\bigskip{}

\centering{}\includegraphics[height=0.08\paperheight]{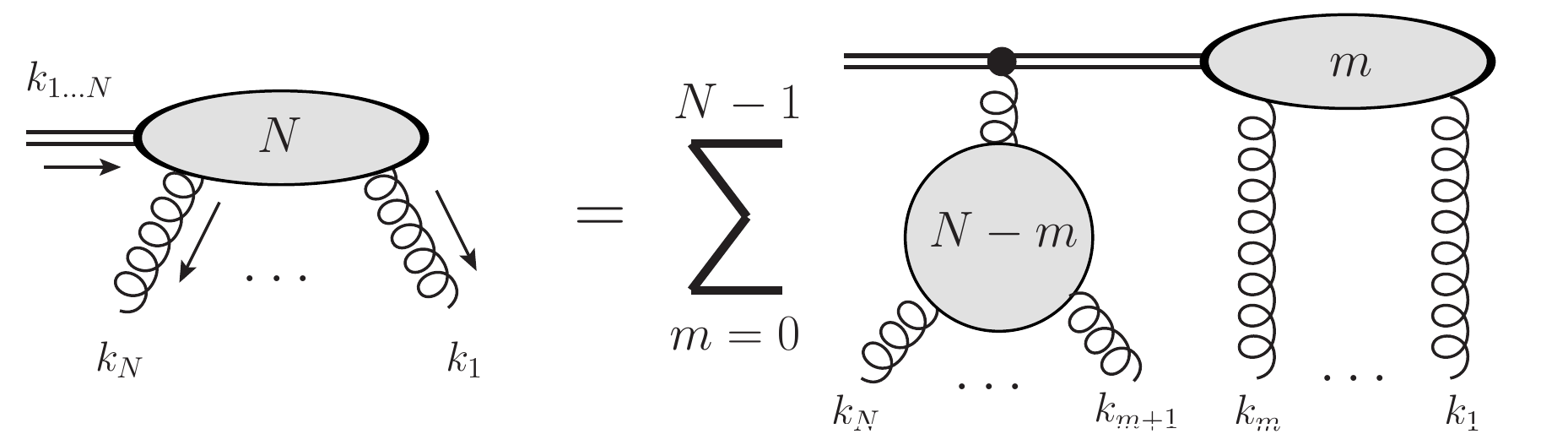} & \centering{}(\myref)\label{eq:Gaugelink_expansion1}\tabularnewline
\end{tabular}

\noindent We will utilize this recursion throughout the rest of the paper; therefore, we will need its algebraic form:
\begin{equation}
\tilde{\mathcal{M}}_{\mathfrak{e}_{1\dots N}}^{\left(\lambda_{1}\dots\lambda_{N}\right)}\left(k_{1\dots N}\right)=\sum_{m=0}^{N-1}\tilde{\mathcal{M}}_{\mathfrak{e}_{1\dots N}}^{\left(\lambda_{1}\dots\lambda_{m}\right)}\left(k_{1\dots m}\right)
\frac{1}{k_{1\dots m}\cdot\varepsilon_{1\dots N}}\,\overline{\mathcal{M}}^{\left(\mathfrak{e}_{1\dots N}\rightarrow\lambda_{m+1}\dots\lambda_{N}\right)}\left(k_{m+1,\dots,N}\right)\, ,\label{eq:Gaugelink_RecRelAlg}
\end{equation}
with
\begin{gather}
\tilde{\mathcal{M}}_{\mathfrak{e}_{1\dots N}}^{\left(\lambda_{i}\right)}\left(k_{i}\right)=i^2 g\, \mathfrak{e}_{1\dots N}^{*}\cdot\varepsilon_{i}^{\lambda_{i}} \, ,\label{eq:Mtildkiki}
\end{gather}
and
$\tilde{\mathcal{M}}_{\mathfrak{e}_{1\dots N}}^{\left(\lambda_{1}\dots\lambda_{m}\right)}\left(k_{1\dots m}\right)=1$ for  $m=0$.
Let us emphasize an important difference between the amplitudes  $\tilde{\mathcal{M}}$ which appear on both sides of  Eq.~(\ref{eq:Gaugelink_RecRelAlg}). The amplitude on the left hand side of ~(\ref{eq:Gaugelink_RecRelAlg}) satisfies property (\ref{eq:Mtild_1}) and thus is gauge invariant. On the contrary, the amplitude $\tilde{\mathcal{M}}$ which appears on the right
hand side of~(\ref{eq:Gaugelink_RecRelAlg})  is not gauge invariant. This stems from the fact that the replacement $\varepsilon_i\leftrightarrow k_i$ will lead to the non-vanishing result
\begin{equation}
\tilde{\mathcal{M}}_{\mathfrak{e}_{1\dots N}}^{\left(\lambda_{1}\dots k_{i}\dots\lambda_{m}\right)}\left(k_{1\dots m}\right)\neq0\,\,\,\,\mathrm{for}\,\,\, i=1,\ldots,m\, .\label{eq:Mtild1mWI}
\end{equation}
This is because the Wilson line slope defining $\tilde{\mathcal{M}}$ is not perpendicular to the off-shell momentum, $\mathfrak{e}_{1\dots N}\cdot k_{1\dots m}\neq0$, as required by (\ref{eq:Mtild_1}).

\subsection{Light-front recursion relation from Wilson lines}

\label{subsec:reltoLF} 

We will now relate the recursion with Wilson lines (\ref{eq:Gaugelink_RecRelAlg}) to the recursion (\ref{eq:masterrecursion}) obtained within the light-front formalism. To this end, we first have to choose the appropriate ``polarization'' vector $\mathfrak{e}_{1\dots N}$ for the off-shell gluon. We choose, of course, the same vector as in the formalism to obtain (\ref{eq:masterrecursion}), i.e. we
choose 
\begin{equation}
\mathfrak{e}_{1\dots N}^{\mu}=\varepsilon_{1\dots N}^{+\mu},
\end{equation}
where $\varepsilon_{1\dots N}^{+}$ is defined by (\ref{eq:LFpolvec1}).
 Note that $\varepsilon_{1\dots N}\cdot k_{1\dots N}=0$, despite that
$k_{1\dots N}$ is off-shell and, thus, $\tilde{\mathcal{M}}_{\varepsilon_{1\dots N}^{+}}^{\left(\lambda_{1}\dots\lambda_{N}\right)}\left(k_{1\dots N}\right)$
is gauge invariant. 
Choosing helicities as $\lambda_{1}=-$, $\lambda_{2}=\dots=\lambda_{N}=+$
and the reference momenta for the polarization vectors to be $\eta$
as in (\ref{eq:LFpolvec1}), we can write (\ref{eq:Gaugelink_RecRelAlg})
as 
\begin{multline}
\tilde{\mathcal{M}}_{\varepsilon_{1\dots N}^{+}}^{\left(-+\dots+\right)}\left(k_{1\dots N}\right)=\overline{\mathcal{M}}^{\left(+\rightarrow-\dots+\right)}\left(k_{1,\dots,N}\right)\\
+\sum_{m=2}^{N-1}\tilde{\mathcal{M}}_{\varepsilon_{1\dots N}^{+}}^{\left(-+\dots+\right)}\left(k_{1\dots m}\right)\frac{1}{\tilde{v}_{\left(1\dots m\right)\left(1\dots N\right)}}\overline{\mathcal{M}}^{\left(+\rightarrow+\dots+\right)}\left(k_{m+1,\dots,N}\right).\label{eq:Gaugelink_RecRelAlg-1}
\end{multline}
Note that now the sum starts with the index $m=2$, as for $m=1$
the term vanishes due to (\ref{eq:Mtildkiki}) and $\varepsilon_{1\dots N}^{+*}\left(\eta\right)\cdot\varepsilon_{1}^{-}\left(\eta\right)=0$
according to (\ref{eq:poldef_2a}). In order to proceed, we have to
find an explicit expression for $\overline{\mathcal{M}}^{\left(+\rightarrow+\dots+\right)}$.
This can be done using the recursion (\ref{eq:Gaugelink_RecRelAlg})
with $\lambda_{1}=\dots=\lambda_{N}=+$ and observing that
\begin{equation}
\tilde{\mathcal{M}}_{\varepsilon_{1\dots N}^{+}}^{\left(+\dots+\right)}=0.
\end{equation}
The details are given in Appendix \ref{sec:Mbar+}. The result reads
\begin{equation}
\overline{\mathcal{M}}^{\left(+\rightarrow+\dots+\right)}\left(k_{1,\dots,N}\right)=-g^{N}\,\frac{\tilde{v}_{\left(1\dots N\right)1}}{\tilde{v}_{1\left(1\dots N\right)}}\,\frac{1}{\tilde{v}_{N\left(N-1\right)}\dots\tilde{v}_{32}\tilde{v}_{21}}.\label{eq:Mbar+}
\end{equation}
Let us note that this is the same as obtained from the light-front 
approach (\ref{eq:mbarsub}). 
To show this we set $i=1$ in (\ref{eq:mbarsub}), and use the relation (\ref{eq:MbarQCDdef}) with  $k_{1\dots N}^2=z_{1\dots N}D_{1\dots N}$. However, even with this encouraging result, the recursion (\ref{eq:Gaugelink_RecRelAlg-1}) is different then the one obtained within the light-front formulation. Indeed, this recursion does not involve the same object on the l.h.s and r.h.s of the equation, as we already discussed below equation~(\ref{eq:Gaugelink_RecRelAlg}).
The recursion relation (\ref{eq:Gaugelink_RecRelAlg-1}) can be, however, written
 entirely in terms
of gauge invariant off-shell amplitudes. That is, we will look for a kernel
$K_{mN}$ such that 
\begin{multline}
\tilde{\mathcal{M}}_{\varepsilon_{1\dots N}^{+}}^{\left(-+\dots+\right)}\left(k_{1\dots N}\right)=\overline{\mathcal{M}}^{\left(+\rightarrow-\dots+\right)}\left(k_{1,\dots,N}\right)\\
+\sum_{m=2}^{N-1}\tilde{\mathcal{M}}_{\varepsilon_{1\dots m}^{+}}^{\left(-+\dots+\right)}\left(k_{1\dots m}\right)\, K_{mN}\,\,\overline{\mathcal{M}}^{\left(+\rightarrow+\dots+\right)}\left(k_{m+1,\dots,N}\right).\label{eq:Gaugelink_RecRelAlg-2}
\end{multline}
Let us underline the difference with relation (\ref{eq:Gaugelink_RecRelAlg-1}).
Now, on the r.h.s., we encounter $\tilde{\mathcal{M}}_{\varepsilon_{1\dots m}^{+}}^{\left(-+\dots+\right)}\left(k_{1\dots m}\right)$,
which does satisfy the Ward identities since $\varepsilon_{1\dots m}^{+}\cdot k_{1\dots m}=0$.
On the contrary, in (\ref{eq:Gaugelink_RecRelAlg-1}) we had $\tilde{\mathcal{M}}_{\varepsilon_{1\dots N}^{+}}^{\left(-+\dots+\right)}\left(k_{1\dots m}\right)$ (note the different Wilson line slope here),  
which is not gauge invariant, as discussed below equation~(\ref{eq:Gaugelink_RecRelAlg}). 
There is one more comment in order here. Our assumption of the existence of the kernel $K_{mN}$ is guided by the light-front result discussed in Sec.~\ref{sec:RecRelation}. It is not obvious however if such kernel exists for any helicity configuration.

In order to find $K_{mN}$, we first
have to find the relation between $\tilde{\mathcal{M}}_{\varepsilon_{1\dots N}^{+}}^{\left(-+\dots+\right)}\left(k_{1\dots m}\right)$
and $\tilde{\mathcal{M}}_{\varepsilon_{1\dots m}^{+}}^{\left(-+\dots+\right)}\left(k_{1\dots m}\right)$.
It can be done by observing that the first term on the r.h.s of (\ref{eq:Gaugelink_RecRelAlg-1})
does not depend on the Wilson line direction (this is the consequence
of the gauge we are using). Thus, we can write this equation separately
for $\tilde{\mathcal{M}}_{\varepsilon_{1\dots N}^{+}}^{\left(-+\dots+\right)}\left(k_{1\dots m}\right)$
and $\tilde{\mathcal{M}}_{\varepsilon_{1\dots m}^{+}}^{\left(-+\dots+\right)}\left(k_{1\dots m}\right)$
and subtract them. Doing this recursively one can find the desired relation. The technical details are given in Appendix~\ref{sec:App_Ginvnoninv}.
Here, we give only the final answer:
\begin{multline}
\tilde{\mathcal{M}}_{\varepsilon_{1\dots N}^{+}}^{\left(-+\dots+\right)}\left(k_{1\dots m}\right)=\tilde{\mathcal{M}}_{\varepsilon_{1\dots m}^{+}}^{\left(-+\dots+\right)}\left(k_{1\dots m}\right)
+\sum_{p=1}^{m-2}\sum_{i_{1}=2}^{m-1}\sum_{i_{2}=2}^{i_{1}-1}\dots\sum_{i_{p}=2}^{i_{p-1}-1}\tilde{\mathcal{M}}_{\varepsilon_{1\dots i_{p}}^{+}}^{\left(-+\dots+\right)}\left(k_{1\dots i_{p}}\right)\\
\qquad\qquad\qquad\qquad\qquad\qquad\overline{\mathcal{M}}^{\left(+\rightarrow+\dots+\right)}\left(k_{i_{p}+1,\dots,i_{p-1}}\right)\dots\overline{\mathcal{M}}^{\left(+\rightarrow+\dots+\right)}\left(k_{i_{1}+1,\dots,m}\right)\\
\frac{\tilde{v}_{\left(1\dots m\right)\left(1\dots N\right)}}{\tilde{v}_{\left(1\dots i_{p}\right)\left(1\dots N\right)}\tilde{v}_{\left(1\dots m\right)\left(1\dots i_{p}\right)}}\,\frac{1}{\tilde{v}_{\left(1\dots i_{1}\right)\left(1\dots i_{p}\right)}\dots\tilde{v}_{\left(1\dots i_{p-1}\right)\left(1\dots i_{p}\right)}}.\label{eq:Ginvnoninv}
\end{multline}
The desired relation is obtained by inserting the above formula into Eq.~(\ref{eq:Gaugelink_RecRelAlg-1}).
However, the result has a very complicated structure containing a tower of sums.
Remarkably, it turns out, that these sums satisfy an equation which resembles
an old-fashioned propagator theory and the solution to this equation
can be found. We relegate all the technical details to Appendix~\ref{sec:App_RecProof} and here we restrict ourselves 
to a pictorial description. 
Namely, the kernel $K_{mN}$ can be thought of as a propagator for certain Hamiltonian which -- after algebraic manipulations -- is expressed by the tower of sums and the expression with $\tilde{v}_{ij}$ in (\ref{eq:Ginvnoninv}). A careful inspection of the sums reveals that they are  ordered in a way that resembles time ordering of old fashioned perturbation theory. One can introduce then  an object that plays the role of the free propagator (see Eq.~(\ref{eq:App_ztild}) in the appendix) and  another one which can be interpreted as a vertex (Eq.~(\ref{eq:App_hidef}) in the appendix). It becomes then clear that all the sums, except one, form again the full propagator, see diagram (\ref{eq:App_prop2}). One can then write the compact integral equation for the kernel $K_{mN}$. Finally it is easy to show that its solution gives the desired kernel with the simple form: 
\begin{equation}
K_{mN}=\frac{z_{1\dots N}}{z_{m+1\dots N}\tilde{v}_{\left(1\dots m+1\right)\left(m+1\right)}}.\label{eq:GI_KmN}
\end{equation}

It is straightforward to check that (\ref{eq:Gaugelink_RecRelAlg-2}) with (\ref{eq:GI_KmN}) and (\ref{eq:Mbar+})
coincides exactly with (\ref{eq:masterrecursion}) obtained within the light-front approach. To this end one only needs 
to redefine $\mathcal{M}$ and $M$ in (\ref{eq:masterrecursion}) to include the energy denominators with appropriate $z_i$'s forming in fact propagators. 
This means also, that the MHV off-shell amplitude $M$ in (\ref{eq:masterrecursion})
is indeed gauge invariant since it corresponds to the gauge invariant $\tilde{\mathcal{M}}$
from the Wilson line approach. Yet another confirmation of this result comes from
the work \cite{vanHameren:2014iua} where similar off-shell gauge invariant
helicity amplitude was calculated and turned out to have also the MHV form.

\section{Conclusions}

\label{sec:Conclusions}

In this work we have analyzed  gauge invariance properties of the gluon off-shell scattering amplitudes with Maximal Helicity Violating configurations using the  light-front formalism. The recurrence relation for such amplitudes, that was first derived in \cite{Cruz-Santiago:2013vta},
encodes a new object, which is off-shell but has the form of an on-shell MHV amplitude.
We demonstrate that this new amplitude is a gauge invariant despite its off-shellness. In order to check the gauge invariance within the light-front formalism, we had to find a way to verify the Ward identities. Unlike in the standard formulation of QCD this is not straightforward on the light front, as
the minus light-cone components  are integrated out by default within this formalism and the standard QCD prescription has to be modified.
 The proper treatment of the Ward identities involves additional instantaneous-like terms, which effectively can be taken into account by forcing the full momentum conservation in the vertices.
Furthermore, we recognize that the light-front recurrence relation  has a very similar form to a recurrence that is encoded in the Wilson line formulation of off-shell amplitudes \cite{Kotko:2014aba}. In fact, we prove that they are precisely the same. Therefore, remarkably,  the new off-shell amplitude that appears in the light-front recursion  is gauge invariant.

As far as different helicity configurations are considered, the situation is much more complicated. A recurrence relation, similar to Eq.~(\ref{eq:Gaugelink_RecRelAlg-1}), can be easily written for any helicity configuration using the Wilson line approach, c.f. Eq.~(\ref{eq:Gaugelink_RecRelAlg}). However, it appears extremely cumbersome and it remains unknown whether it would be possible  to cast it into simple, truly recursive form as in Eq.~(\ref{eq:Gaugelink_RecRelAlg-2}).

Let us, finally, conclude that our study once again stresses the importance of gauge invariance in any QCD computations. As we have shown in the current paper,  the complicated resummation of whole classes of light-front diagrams, as performed in Ref.~\cite{Cruz-Santiago:2013vta}, leads precisely to the straight infinite Wilson line, which is a manifestly gauge invariant object.


\section*{Acknowledgments}
We thank Stanley Brodsky, Leszek Motyka and Radu Roiban for discussions. This work was supported in part  by the DOE  grant No. DE-SC-0002145, grant No. DE-FG02-93ER40771  and by the Polish NCN grant DEC-2011/01/B/ST2/03915.  
\appendix

\section{Useful identities}

\label{sec:Identities}

In this supplement we summarize certain relations for the quantities
$\tilde{v}_{ij}$ and $v_{ij}$ defined in Eqs.~(\ref{eq:vijtild}),~(\ref{eq:vij}).

Straight from the definition, we have an antisymmetry property of
$v_{ij}$
\begin{equation}
v_{ij}=-v_{ji}.\label{eq:Ident1}
\end{equation}
For the $\tilde{v}_{ij}$ the exchange of indices gives
\begin{equation}
\tilde{v}_{ij}=-\frac{k_{i}^{+}}{k_{j}^{+}}\,\tilde{v}_{ji}.\label{eq:Ident2}
\end{equation}
Obviously
\begin{equation}
\tilde{v}_{ii}=0\,, \label{eq:Ident2a}
\end{equation}
what comes out of the identity (\ref{eq:Ident2}) or the transversity
of polarization vectors. Finally, we have the following decomposition
relation
\begin{equation}
\tilde{v}_{ij}-\tilde{v}_{il}=\frac{k_{i}^{+}}{k_{l}^{+}}\tilde{v}_{lj}.\label{eq:Ident3}
\end{equation}
The above relations come straight from the definitions and can be
easily proved.

Let us now consider a set of $\tilde{v}_{ij}$ constructed for momenta
$k_{\left(1\dots N\right)},k_{1},\dots,k_{N}$ such that $k_{1\dots N}=k_{1}+k_{2}+\dots+k_{N}$
. Then we have
\begin{equation}
\tilde{v}_{\left(1\dots i\right)\left(1\dots N\right)}=-\tilde{v}_{\left(i+1,\dots N\right)\left(1\dots N\right)}\label{eq:Ident5}
\end{equation}
and so on. Another useful relation reads
\begin{equation}
\tilde{v}_{\left(i\dots N\right)i}=\tilde{v}_{ii}+\tilde{v}_{\left(i+1\dots N\right)i}=\tilde{v}_{\left(i+1\dots N\right)i} \label{eq:Ident5a}
\end{equation}
thanks to property (\ref{eq:Ident2a}).
Moreover, the following summation relations hold
\begin{equation}
\sum_{j=1}^{N}\frac{\tilde{v}_{\left(1\ldots j\right)j}}{\tilde{v}_{j\left(1\dots j\right)}}\,\tilde{v}_{j\left(j+1\right)}=\tilde{v}_{\left(1\dots N\right)N},\label{eq:Ident4}
\end{equation}
\begin{equation}
\sum_{i=m}^{N-1}\frac{\tilde{v}_{\left(i+1\dots N\right)\left(i+1\right)}}{\tilde{v}_{\left(i+1\right)\left(i+1\dots N\right)}}\,\tilde{v}_{\left(i+1\right)i}=\tilde{v}_{\left(m\dots N\right)m}.\label{eq:Ident6}
\end{equation}
They are proven using (\ref{eq:Ident3}). For example, for (\ref{eq:Ident6})
we have 
\begin{multline}
\sum_{i=m}^{N-1}\frac{\tilde{v}_{\left(i+1\dots N\right)\left(i+1\right)}}{\tilde{v}_{\left(i+1\right)\left(i+1\dots N\right)}}\,\tilde{v}_{\left(i+1\right)i}=-\sum_{i=m}^{N-1}\frac{z_{i+1\dots N}}{z_{i+1}}\,\tilde{v}_{\left(i+1\right)i}\\
=\sum_{i=m}^{N-1}\left(\tilde{v}_{\left(i+1\dots N\right)\left(i+1\right)}-\tilde{v}_{\left(i+1\dots N\right)i}\right)=\sum_{k=m+1}^{N}\tilde{v}_{\left(k\dots N\right)k}-\sum_{i=m}^{N-1}\tilde{v}_{\left(i\dots N\right)i}\\
=-\tilde{v}_{\left(m\dots N\right)m}+\tilde{v}_{NN}=-\tilde{v}_{\left(m\dots N\right)m}
\end{multline}
Above, we have used (\ref{eq:Ident2}) and \ref{eq:Ident5a}.

\section{Explicit example for the Ward identity check on the light-front}

\label{sec:App_WI}

In this appendix we will demonstrate that, when checking the Ward
identity within the light-front formalism, additional instantaneous-like terms
appear. These terms, in fact, recover the full momentum conservation.

We will use a specific, yet quite general example. Consider the following diagram

\begin{tabular}{>{\centering}m{0.87\columnwidth}>{\centering}m{0.05\columnwidth}}
\bigskip{}

\begin{centering}
\parbox[c]{0.18\textwidth}{$\mathcal{M}^{\lambda_2}\left(P,k_{2},k_{3}\right)=$}\parbox[c]{0.36\textwidth}{\includegraphics[width=0.35\textwidth]{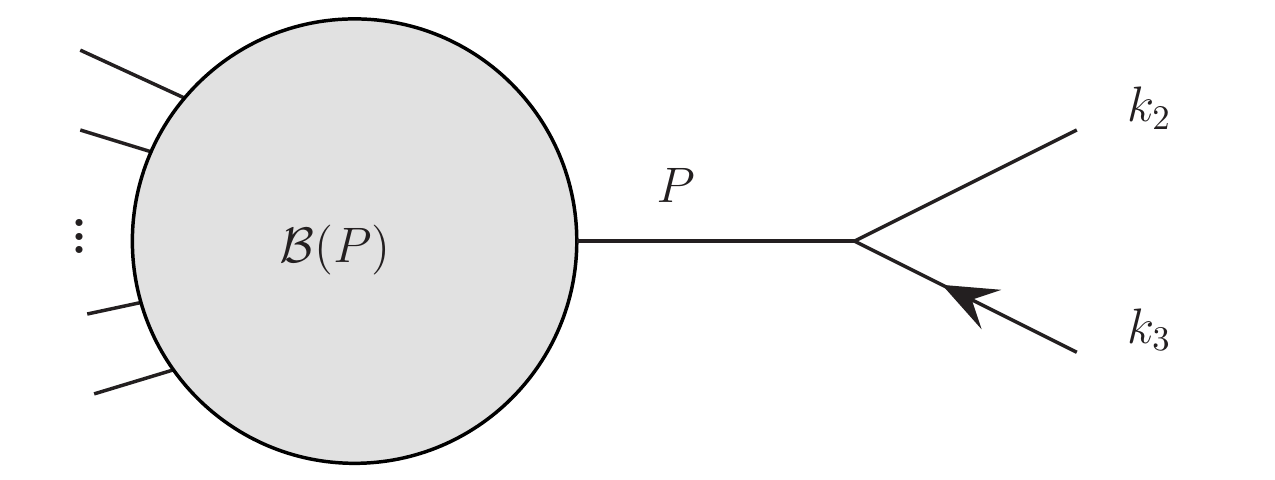}}
\par\end{centering}

\smallskip{}
 & \centering{}(\myref)\label{eq:App_WI1}\tabularnewline
\end{tabular}

\noindent where the arrow denotes the contraction with the momentum instead
of the corresponding polarization vector. The momenta labels are chosen
in such a way that the contact with Fig.~\ref{fig:M1to3WI} can be
established. The amplitude can be expressed as
\begin{equation}
\mathcal{M}^{\lambda_2}\left(P,k_{2},k_{3}\right)=\int\!\! d^{4}x\,\tilde{\mathcal{B}}^{\mu}\left(x;P\right)g_{\mu\nu}\tilde{\mathcal{A}}^{\nu\lambda_{2}}\left(x;k_{2},k_{3}\right),
\end{equation}
where
\begin{equation}
\tilde{\mathcal{B}}^{\mu}\left(x;P\right)=e^{iP\cdot x}\mathcal{B}^{\mu}\left(P\right)
\end{equation}
represents the blob and $\tilde{\mathcal{A}}^{\nu\lambda_{2}}\left(x;k_{2},k_{3}\right)$ represents
the gluon splitting. The 
$\lambda_{2}$ is the polarization of the gluon with momentum
$k_{2}$. The gluon propagator is included in~$\tilde{\mathcal{A}}$.

For the purpose of this example, the transition to the light-front formalism will
be done according to \cite{Kogut:1969xa}. The basic idea is to integrate
the minus components in the propagators. In our example, we
will first reintroduce the suitable integration. In order to proceed
let us decompose the tensor $g_{\mu\nu}$ as follows
\begin{equation}
-g_{\mu\nu}=\sum_{\lambda_{23}=\pm}\varepsilon_{\mu}^{\lambda_{23}}\varepsilon_{\nu}^{\lambda_{23}*}+\dots,
\label{eq:poldecomp}
\end{equation}
where 
$\lambda_{23}$ is the polarization of the intermediate gluon and 
the dots stand for the gauge terms and the term that will lead
to the instantaneous interactions. For the purpose of the present example
we will retain only the part containing the polarization vectors and
show that the instantaneous-like term  appears (in addition to the standard
instantaneous term originating from the terms represented by the dots in (\ref{eq:poldecomp})).
 The corresponding contribution reads
\begin{equation}
\mathcal{M}'\,^{\lambda_2}\left(P,k_{2},k_{3}\right)=\sum_{\lambda_{23}}\int\!\! d^{4}x\,\tilde{\mathcal{B}}^{\lambda_{23}}\left(x;P\right)\tilde{\mathcal{A}}^{\lambda_{23}\lambda_{2}}\left(x;k_{2},k_{3}\right),\label{eq:App_WI_Mprim}
\end{equation}
where
\begin{equation}
\tilde{\mathcal{B}}^{\lambda_{23}}\left(x;P\right)=e^{iP\cdot x}\mathcal{B}^{\mu}\left(P\right)\varepsilon_{\mu}^{\lambda_{23}*}\left(P\right)
\end{equation}
and
\begin{equation}
\mathcal{A}^{\lambda_{23}\lambda_{2}}\left(x;k_{2},k_{3}\right)= e^{-ix\cdot \left(k_2+k_3\right)}
 \frac{-i}{\left(k_{2}+k_{3}\right)^{2}+i\epsilon}\, V_{3}^{\lambda_{23}\lambda_{2}k_{3}}\left(-k_{2}-k_{3},k_{2},k_{3}\right)\label{eq:App_WI_Adef}
\end{equation}
with the generic color-ordered triple gluon vertex 
\begin{multline}
V_{3}^{\lambda_{1}\lambda_{2}\lambda_{3}}\left(p_{1},p_{2},p_{3}\right)=i\varepsilon_{\alpha_{1}}^{\lambda_{1}}\left(p_{1}\right)\varepsilon_{\alpha_{2}}^{\lambda_{2}}\left(p_{2}\right)\varepsilon_{\alpha_{3}}^{\lambda_{3}}\left(p_{3}\right)\\
\left[g^{\alpha_{1}\alpha_{2}}\left(p_{1}-p_{2}\right)^{\alpha_{3}}+g^{\alpha_{2}\alpha_{3}}\left(p_{2}-p_{3}\right)^{\alpha_{1}}+g^{\alpha_{3}\alpha_{1}}\left(p_{3}-p_{1}\right)^{\alpha_{2}}\right].
\end{multline}
The $k_{3}$ instead of a polarization superscript for $V_{3}$ in (\ref{eq:App_WI_Adef}) denotes
the replacement $\varepsilon_{3}\leftrightarrow k_{3}$. We can rewrite
(\ref{eq:App_WI_Adef}) as
\begin{equation}
\tilde{\mathcal{A}}^{\lambda_{23}\lambda_{2}}\left(x;k_{2},k_{3}\right)=\int\!\! d^{4}k_{23}\, e^{-ik_{23}\cdot x}\frac{-i}{k_{23}^{2}+i\epsilon}\, V_{3}^{\lambda_{23}\lambda_{2}k_{3}}\left(-k_{23},k_{2},k_{3}\right)\delta^{4}\left(k_{23}-k_{2}-k_{3}\right).\label{eq:App_WI_Adelt}
\end{equation}
Above, we have restored the unintegrated propagator (the scalar part).
Switching to light-cone variables and 
using the integral representation for
 the Dirac delta for the minus
component, we have 
\begin{multline}
\tilde{\mathcal{A}}^{\lambda_{23}\lambda_{2}}\left(x;k_{2},k_{3}\right)=\frac{1}{2}\int \frac{dy^{+}}{2\pi}\,e^{-i\frac{1}{2}y^{+}\left(k_{2}^{-}+k_{3}^{-}\right)} \\
\int\!\! dk_{23}^{+}\, e^{-i\frac{1}{2}k_{23}^{+}x^{-}}\int\!\! d^{2}k_{23T}\, e^{i\vec{k}_{23T}\cdot\vec{x}_{T}}\int\!\! dk_{23}^{-}\, e^{-i\frac{1}{2}k_{23}^{-}\cdot\left(x^{+}-y^{+}\right)}\\
\frac{-i}{k_{23}^{+}k_{23}^{-}-k_{23T}^{2}+i\epsilon}\, V_{3}^{\lambda_{23}\lambda_{2}k_{3}}\left(-k_{23},k_{2},k_{3}\right)\delta\left(k_{23}^{+}-k_{2}^{+}-k_{3}^{+}\right)\delta^{2}\left(\vec{k}_{23T}-\vec{k}_{2T}-\vec{k}_{3T}\right).\label{eq:App_WI_A1}
\end{multline}
Since the polarization vectors project only on plus and transverse
components (c.f. (\ref{eq:LFpolvec1})) we can write
\begin{equation}
V_{3}^{\lambda_{23}\lambda_{2}k_{3}}\left(-k_{23},k_{2},k_{3}\right)=\mathcal{V}_{1}+\mathcal{V}_{2},\label{eq:App_WI_V3}
\end{equation}
where
\begin{equation}
\mathcal{V}_{1}=-ik_{23}\cdot k_{3}\,\,\varepsilon_{2}^{\lambda_{2}}\cdot\varepsilon_{23}^{\lambda_{23}},
\end{equation}
\begin{equation}
\mathcal{V}_{2}=i-k_{2}\cdot k_{3}\,\,\varepsilon_{2}^{\lambda_{2}}\cdot\varepsilon_{23}^{\lambda_{23}}.
\end{equation}
The contribution $\mathcal{V}_{1}$ contains $k_{23}\cdot k_{3}$ in the
numerator, i.e. the integration variable and thus will lead to instantaneous-like
term (this is not the case for $\mathcal{V}_{2}$). Therefore, in what
follows we consider a contribution to (\ref{eq:App_WI1}) only from
$\mathcal{V}_{1}$. It reads
\begin{multline}
\tilde{\mathcal{A}}_{1}\left(x;k_{2},k_{3}\right)=-\frac{1}{2}\int \frac{dy^{+}}{2\pi}\,e^{-i\frac{1}{2}y^{+}\left(k_{2}^{-}+k_{3}^{-}\right)}\int\!\! dk_{23}^{+}\, e^{-i\frac{1}{2}k_{23}^{+}x^{-}}\int\!\! d^{2}k_{23T}\, e^{i\vec{k}_{23T}\cdot\vec{x}_{T}}\\
\varepsilon^{\lambda_{2}}\left(k_{2}\right)\cdot\varepsilon^{\lambda_{23}}\left(k_{23}\right)\int\!\! dk_{23}^{-}\, e^{-i\frac{1}{2}k_{23}^{-}\cdot\left(x^{+}-y^{+}\right)}\frac{\frac{1}{2}k_{23}^{-}k_{3}^{+}+\frac{1}{2}k_{23}^{+}k_{3}^{-}-\vec{k}_{23T}\cdot\vec{k}_{3T}}{k_{23}^{+}k_{23}^{-}-k_{23T}^{2}+i\epsilon}\,\\
\delta\left(k_{23}^{+}-k_{2}^{+}-k_{3}^{+}\right)\delta^{2}\left(\vec{k}_{23T}-\vec{k}_{2T}-\vec{k}_{3T}\right).\label{eq:App_WI_A1mod}
\end{multline}
The light-front approach is achieved by integrating over the $k_{23}^{-}$ component.
The relevant integral can be done by the residue technique and reads (since $k_2^+,k_3^+>0$ only $k_{23}^+>0$ part gives contribution)
\begin{multline}
\Theta\left(k_{23}^{+}\right)\int \frac{dk_{23}^{-}}{2\pi}\,\, e^{-i\frac{1}{2}k_{23}^{-}\cdot\left(x^{+}-y^{+}\right)}\frac{\frac{1}{2}k_{23}^{-}k_{3}^{+}+\frac{1}{2}k_{23}^{+}k_{3}^{-}-\vec{k}_{23T}\cdot\vec{k}_{3T}}{k_{23}^{+}k_{23}^{-}-k_{23T}^{2}+i\epsilon}=\frac{\Theta\left(k_{23}^{+}\right)}{k_{23}^{+}}\,\\
\left[-i\hat{k}_{23}\cdot k_{3}\Theta\left(x^{+}-y^{+}\right)e^{-i\frac{1}{2}\hat{k}_{23}^{-}\left(x^{+}-y^{+}\right)}+\frac{1}{2}\delta\left(x^{+}-y^{+}\right)k_{3}^{+}\right]\label{eq:integ2}
\end{multline}
where
\begin{equation}
\hat{k}_{23}\cdot k_{3}=\frac{1}{2}\hat{k}_{23}^{-}k_{3}^{+}+\frac{1}{2}k_{23}^{+}k_{3}^{-}-\vec{k}_{23T}\cdot\vec{k}_{3T}
\end{equation}
with
\begin{equation}
\hat{k}_{23}^{-}=\frac{k_{23T}^{2}}{k_{23}^{+}}
\end{equation}
is the scalar product with the minus component of $k_{23}$
set to the on-shell value, as given by the residue. Note that the
first term of the r.h.s. of (\ref{eq:integ2}) is the one that was
taken into account in the example (\ref{eq:M1to3WI2}) and alone leads to an
incorrect result. Clearly, the second term of (\ref{eq:integ2}) was
missing. Let us now calculate the contribution to the amplitude $\mathcal{M}'$
coming from the first term in (\ref{eq:integ2}). Performing the integrals over the light-cone time we get
\begin{multline}
\mathcal{M}_{1a}'^{\,\lambda_2}\left(P,k_{2},k_{3}\right)=-\frac{2}{P^{+}}\sum_{\lambda_{23}}\,
\left(2\pi\right)^4
\delta\left(P^{+}-k_{2}^{+}-k_{3}^{+}\right)\delta\left(P^{-}-k_{2}^{-}-k_{3}^{-}\right)\delta^{2}\left(\vec{P}_{T}-\vec{k}_{2T}-\vec{k}_{3T}\right)\\
\Theta\left(P^{+}\right)\frac{1}{P^{-}-\hat{k}_{23}^{-}+i\epsilon}\,\, \mathcal{B}^{\lambda_{23}}\left(P\right)\,\,
\hat{k}_{23}\cdot k_{3}\,\,\,
\varepsilon^{\lambda_{2}}\left(k_{2}\right)\cdot\varepsilon^{\lambda_{23}}
\left(k_{23}\right).
\end{multline}
Above (and below), $\hat{k}_{23}^-$ is understood as 
\begin{equation}
\hat{k}_{23}^- = \frac{\left(\vec{k}_{2T}+\vec{k}_{3T}\right)^2}{k_2^+ + k_3^+}\, .
\end{equation}
Next, the contribution of the instantaneous-like term reads
\begin{multline}
\mathcal{M}_{1b}'^{\,\lambda_2}\left(P,k_{2},k_{3}\right)=-\frac{k_{3}^{+}}{2P^{+}}\sum_{\lambda_{23}}\,
\left(2\pi\right)^4
\delta\left(P^{+}-k_{2}^{+}-k_{3}^{+}\right)
\delta\left(P^{-}-k_{2}^{-}-k_{3}^{-}\right)\delta^{2}\left(\vec{P}_{T}-\vec{k}_{2T}-\vec{k}_{3T}\right)\,
\\
\Theta\left(P^{+}\right)\mathcal{B}^{\lambda_{23}}\left(P\right)\,\,\varepsilon^{\lambda_{2}}
\left(k_{2}\right)\cdot\varepsilon^{\lambda_{23}}\left(k_{23}\right).
\end{multline}
Both contributions have to be added:
\begin{multline}
\mathcal{M}_{1a}'^{\,\lambda_2}\left(P,k_{2},k_{3}\right)+\mathcal{M}_{1b}'^{\,\lambda_2}\left(P,k_{2},k_{3}\right)\\
=-\frac{2}{P^{+}}\sum_{\lambda_{23}}\,
^{\,\lambda_2}
\delta\left(P^{+}-k_{2}^{+}-k_{3}^{+}\right)\delta\left(P^{-}-k_{2}^{-}-k_{3}^{-}\right)\delta^{2}\left(\vec{P}_{T}-\vec{k}_{2T}-\vec{k}_{3T}\right)\\
\Theta\left(P^{+}\right)\mathcal{B}^{\lambda_{23}}\left(P\right)\,\,\varepsilon^{\lambda_{2}}\left(k_{2}\right)\cdot\varepsilon^{\lambda_{23}}\left(k_{23}\right)
\left[\frac{\hat{k}_{23}\cdot k_{3}}{P^{-}-\hat{k}_{23}^{-}+i\epsilon}+\frac{1}{2}k_{3}^{+}\right]\label{eq:App_WI_M1sum}
\end{multline}
However, the square bracket can be rewritten as 
\begin{equation}
\frac{\hat{k}_{23}\cdot k_{3}}{P^{-}-\hat{k}_{23}^{-}+i\epsilon}+\frac{1}{2}k_{3}^{+}=\frac{k_{23}\cdot k_{3}}{P^{-}-\hat{k}_{23}^{-}+i\epsilon},
\end{equation}
where we have used the delta function $\delta\left(P^{-}-k_{2}^{-}-k_{3}^{-}\right)$
appearing in (\ref{eq:App_WI_M1sum}) to write the scalar product on the r.h.s.
This is indeed the correct contribution from the vertex (\ref{eq:App_WI_V3})
as is easily seen by integrating back the delta $\delta^{4}\left(k_{23}-k_{2}-k_{3}\right)$
in (\ref{eq:App_WI_Adelt}).

\section{Off-shell $+\rightarrow+\ldots+$ amplitude from Wilson lines}

\label{sec:Mbar+}

Let us consider the gauge invariant amplitude $\tilde{\mathcal{M}}_{+}^{\left(+\dots+\right)}$
with the choice of the polarization vectors~(\ref{eq:LFpolvec1}).
As mentioned in Sec.~\ref{sec:Notation} this corresponds to choosing
$\eta$ as the reference momentum for all the polarization vectors.
Since $\tilde{\mathcal{M}}_{+}^{\left(+\dots+\right)}\equiv\tilde{\mathcal{M}}_{\varepsilon_{1\dots N}^{+}\left(\eta\right)}^{\left(\varepsilon_{1}^{+}\left(\eta\right)\dots\varepsilon_{N}^{+}\left(\eta\right)\right)}$
is gauge invariant, we can freely change the reference momenta of the
polarization vectors $\varepsilon_{1}^{+}\left(\eta\right),\dots,\varepsilon_{N}^{+}\left(\eta\right)$.
Let us thus use (\ref{eq:LFpolvec3}) and set $k_{1\dots N}$ as the
reference momentum
\begin{equation}
\tilde{\mathcal{M}}_{\varepsilon_{1\dots N}^{+}\left(\eta\right)}^{\left(\varepsilon_{1}^{+}\left(\eta\right)\dots\varepsilon_{N}^{+}\left(\eta\right)\right)}=\tilde{\mathcal{M}}_{\varepsilon_{1\dots N}^{+}\left(\eta\right)}^{\left(\varepsilon_{1}^{+}\left(k_{1\dots N}\right)\dots\varepsilon_{N}^{+}\left(k_{1\dots N}\right)\right)}.\label{eq:App_Mplus1}
\end{equation}
Note that the properties (\ref{eq:poldef_1a})-(\ref{eq:poldef_2a})
still hold for $\varepsilon_{i}\left(k_{1\dots N}\right)$ despite
the fact that $k_{1\dots N}$ is off-shell (actually, only the last
relation of (\ref{eq:poldef_2a}) is non-trivial to check). The amplitude
(\ref{eq:App_Mplus1}) is given by the expansion (\ref{eq:Gaugelink_expansion}).
Consider any blob  $\overline{\mathcal{M}}^{\left(\varepsilon_{1\dots N}^{+}\left(\eta\right)\rightarrow\varepsilon_{i}^{+}\left(k_{1\dots N}\right)\dots\varepsilon_{j}^{+}\left(k_{1\dots N}\right)\right)}$, 
$j>i$, attached to the Wilson line. Such blob contains terms with at least one scalar product
of polarization vectors. This is due to the following standard argument
(see e.g. \cite{Mangano:1990by})). Since the are at most $j-i-1$
triple gluon vertices there may be at most $j-i-1$ momentum vectors
in the numerator. These vectors are contracted with $j-i+1$ polarization
vectors, which means that at least two polarization vectors must be
contracted together. Due to our choice of reference momenta all such
scalar products vanish due to (\ref{eq:poldef_2a}). This happens
for all the blobs, therefore
\begin{equation}
\tilde{\mathcal{M}}_{+}^{\left(+\dots+\right)}=0.\label{eq:App_Mplus2}
\end{equation}
Of course, for the reference momenta set to $\eta$ the blobs itself
no longer vanish, but different contributions get cancelled due to the
gauge invariance. 

Let us now look at the consequences of the above equation. Consider
the recursion (\ref{eq:Gaugelink_expansion1}) for $N=2$ and the Wilson line slope set to a vector
$\mathfrak{u}$ defined by 
\begin{equation}
\mathfrak{u}^{\mu}=\varepsilon_{\perp}^{+\mu}+\frac{\vec{\varepsilon}_{\perp}^{\,\,+}\cdot\vec{p}_{\perp}}{p\cdot\eta}\,\eta^{\mu}\label{eq:App_Mplus0}
\end{equation}
for certain momentum $p$ (for example for $p=k_{1\dots N}$ we have $\mathfrak{u}=\varepsilon_{1\dots N}$, but we want to keep it more general here). We have
\begin{equation}
\tilde{\mathcal{M}}_{\mathfrak{u}}^{\left(++\right)}\left(k_{12}\right)=
\overline{\mathcal{M}}^{\left(\mathfrak{u}\rightarrow++\right)}\left(k_{12}\right)
+\tilde{\mathcal{M}}_{\mathfrak{u}}^{\left(+\right)}\left(k_{1}\right)\frac{1}{k_{1}\cdot\mathfrak{u}}\overline{\mathcal{M}}^{\left(\mathfrak{u}\rightarrow+\right)}\left(k_{2}\right).\label{eq:App_Mplus5}
\end{equation}
If $\mathfrak{u}=\varepsilon_{12}^{+}$ the l.h.s vanishes according
to (\ref{eq:App_Mplus2}) and we have 
\begin{equation}
\overline{\mathcal{M}}^{\left(+\rightarrow++\right)}\left(k_{12}\right)=
-\tilde{\mathcal{M}}_{+}^{\left(+\right)}\left(k_{1}\right)\frac{1}{k_{1}\cdot\varepsilon_{12}}\overline{\mathcal{M}}^{\left(+\rightarrow+\right)}\left(k_{2}\right)\label{eq:App_Mplus6}
\end{equation}
or diagrammatically 

\begin{tabular}{>{\centering}m{0.83\columnwidth}>{\centering}m{0.05\columnwidth}}
\bigskip{}

\centering{}\includegraphics[height=0.06\paperheight]{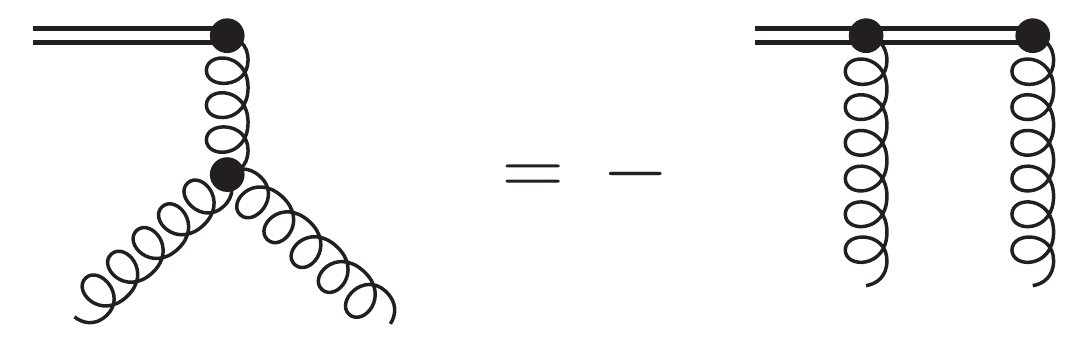} & \centering{}(\myref)\label{eq:App_Mplus3}\tabularnewline
\end{tabular}

\noindent Calculating the r.h.s we get  (remember that by convention we include an additional $i$ factor, c.f. (\ref{eq:MbarQCDdef}))
\begin{equation}
\overline{\mathcal{M}}^{\left(+\rightarrow++\right)}\left(k_{12}\right)=-i^2\left(ig\right)^{2}\,\frac{\varepsilon_{12}^{-}\cdot\varepsilon_{1}^{+}\,\varepsilon_{12}^{-}\cdot\varepsilon_{2}^{+}}{k_{1}\cdot\varepsilon_{12}^{-}}=-g^{2}\,\frac{1}{\tilde{v}_{1\left(12\right)}}=g^{2}\frac{z_{12}}{z_{1}}\,\frac{1}{\tilde{v}_{21}}.\label{eq:App_Mplus4}
\end{equation}
Inserting this back to (\ref{eq:App_Mplus5}) we get
\begin{equation}
\tilde{\mathcal{M}}_{\mathfrak{u}}^{\left(++\right)}\left(k_{12}\right)=-g^{2}\,\left(\frac{1}{\tilde{v}_{1\left(12\right)}}-\frac{1}{\tilde{v}_{1\left(\mathfrak{u}\right)}}\right)=g^{2}\,\frac{z_{1}}{z_{\mathfrak{u}}}\,\frac{\tilde{v}_{\left(\mathfrak{u}\right)\left(12\right)}}{\tilde{v}_{1\left(12\right)}\tilde{v}_{1\left(\mathfrak{u}\right)}},\label{eq:App_Mplus7}
\end{equation}
where we have used (\ref{eq:Ident3}) and introduced
\begin{equation}
\tilde{v}_{i(\mathfrak{u})} = k_i\cdot \mathfrak{u},\quad z_{\mathfrak{u}}=p^+\,.
\end{equation}
 Also, we have utilized the
fact that in the light-cone gauge
\begin{equation}
\overline{\mathcal{M}}^{\left(+\rightarrow++\right)}\left(k_{12}\right)=
\overline{\mathcal{M}}^{\left(\mathfrak{u}\rightarrow++\right)}\left(k_{12}\right),\label{eq:App_Mplus8}
\end{equation}
as the propagator on the l.h.s of (\ref{eq:App_Mplus3}) always projects
(\ref{eq:App_Mplus0}) to $\varepsilon_{12}^{+}$.

For $N=3$ we have
\begin{multline}
\tilde{\mathcal{M}}_{\mathfrak{u}}^{\left(+++\right)}\left(k_{123}\right)
=\overline{\mathcal{M}}^{\left(\mathfrak{u}\rightarrow+++\right)}\left(k_{123}\right)
+\tilde{\mathcal{M}}_{\mathfrak{u}}^{\left(+\right)}\left(k_{1}\right)\frac{1}{k_{1}\cdot\mathfrak{u}}\overline{\mathcal{M}}^{\left(\mathfrak{u}\rightarrow++\right)}
\left(k_{23}\right)\\
+\tilde{\mathcal{M}}_{\mathfrak{u}}^{\left(++\right)}\left(k_{12}\right)\frac{1}{k_{12}\cdot\mathfrak{u}}\overline{\mathcal{M}}^{\left(\mathfrak{u}\rightarrow+\right)}\left(k_{3}\right)\label{eq:App_Mplus9}
\end{multline}
Inserting (\ref{eq:App_Mplus7}) we get
\begin{equation}
\tilde{\mathcal{M}}_{\mathfrak{u}}^{\left(+++\right)}\left(k_{123}\right)=
\overline{\mathcal{M}}^{\left(\mathfrak{u}\rightarrow+++\right)}\left(k_{123}\right)\\
-g^{3}\left(\frac{1}{\tilde{v}_{1\left(\mathfrak{u}\right)}\tilde{v}_{2\left(23\right)}}-\frac{z_{1}}{z_{\mathfrak{u}}}\,\frac{\tilde{v}_{\left(\mathfrak{u}\right)\left(12\right)}}{\tilde{v}_{1\left(12\right)}\tilde{v}_{1\left(\mathfrak{u}\right)}\tilde{v}_{\left(12\right)\left(\mathfrak{u}\right)}}\right).\label{eq:App_Mplus10}
\end{equation}
Setting $\mathfrak{u}=\varepsilon_{123}^{+}$ we eliminate l.h.s and
thus
\begin{equation}
\overline{\mathcal{M}}^{\left(+\rightarrow+++\right)}\left(k_{123}\right)=g^{3}\left(\frac{1}{\tilde{v}_{1\left(123\right)}\tilde{v}_{2\left(23\right)}}-\frac{z_{1}}{z_{123}}\,\frac{\tilde{v}_{\left(123\right)\left(12\right)}}{\tilde{v}_{1\left(12\right)}\tilde{v}_{1\left(123\right)}\tilde{v}_{\left(12\right)\left(123\right)}}\right)
=-g^{3}\,\frac{\tilde{v}_{\left(123\right)1}}{\tilde{v}_{1\left(123\right)}}\frac{1}{\tilde{v}_{32}\tilde{v}_{21}}.\label{eq:App_Mplus11}
\end{equation}
Again, one can calculate $\tilde{\mathcal{M}}_{\mathfrak{u}}^{\left(+++\right)}$
by inserting the above to (\ref{eq:App_Mplus10}).

The above results generalize. We have
\begin{equation}
\overline{\mathcal{M}}^{\left(+\rightarrow+\dots+\right)}\left(k_{1\dots N}\right)=-g^{N}\,\frac{\tilde{v}_{\left(1\dots N\right)1}}{\tilde{v}_{1\left(1\dots N\right)}}\,\frac{1}{\tilde{v}_{N\left(N-1\right)}\dots\tilde{v}_{32}\tilde{v}_{21}}\label{eq:App_Mplus12}
\end{equation}
and thus
\begin{multline}
\tilde{\mathcal{M}}_{\mathfrak{u}}^{\left(+\dots+\right)}\left(k_{1\dots N}\right)=-g^{N}\,\left(\frac{1}{\tilde{v}_{1\left(1\dots N\right)}}-\frac{1}{\tilde{v}_{1\left(\mathfrak{u}\right)}}\right)\,\frac{\tilde{v}_{\left(1\dots N\right)1}}{\tilde{v}_{N\left(N-1\right)}\dots\tilde{v}_{32}\tilde{v}_{21}}\\
=\left(-g\right)^{N}\,\frac{\tilde{v}_{\left(1\dots N\right)\left(\mathfrak{u}\right)}}{\tilde{v}_{1\left(\mathfrak{u}\right)}}\frac{1}{\tilde{v}_{N\left(N-1\right)}\dots\tilde{v}_{32}\tilde{v}_{21}}\label{eq:App_Mplus13}
\end{multline}
The proof is by checking, that these expressions
satisfy the recursion relation (\ref{eq:Gaugelink_expansion1}) rewritten
for the current helicity case and for $\mathfrak{u}=\varepsilon_{1\dots N}^{+}$. That is, we need to verify if
\begin{equation}
\overline{\mathcal{M}}^{\left(+\rightarrow+\dots+\right)}\left(k_{1,\dots,N}\right)=-\sum_{m=1}^{N-1}\tilde{\mathcal{M}}_{\varepsilon_{1\dots N}^{+}}^{\left(+\dots+\right)}\left(k_{1\dots m}\right)\overline{\mathcal{M}}^{\left(+\rightarrow+\dots+\right)}\left(k_{m+1,\dots,N}\right)\frac{1}{\tilde{v}_{\left(1\dots m\right)\left(1\dots N\right)}}.\label{eq:App_Mplus14}
\end{equation}
The r.h.s. with (\ref{eq:App_Mplus12}) and (\ref{eq:App_Mplus13})
reads
\begin{multline}
-\sum_{m=1}^{N-1}g^{m}\,\frac{\tilde{v}_{\left(1\dots m\right)\left(1\dots N\right)}}{\tilde{v}_{1\left(1\dots N\right)}}\,\frac{1}{\tilde{v}_{m\left(m-1\right)}\dots\tilde{v}_{32}\tilde{v}_{21}}\,\\
\left[g^{N-m}\,\frac{\tilde{v}_{\left(m+1\dots N\right)\left(m+1\right)}}{\tilde{v}_{\left(m+1\right)\left(m+1\dots N\right)}}\,\frac{1}{\tilde{v}_{N\left(N-1\right)}\dots\tilde{v}_{\left(m+2\right)\left(m+1\right)}}\right]\frac{1}{\tilde{v}_{\left(1\dots m\right)\left(1\dots N\right)}}\\
=-g^{N}\,\sum_{m=1}^{N-1}\,\frac{1}{\tilde{v}_{1\left(1\dots N\right)}}\,\frac{\tilde{v}_{\left(m+1\dots N\right)\left(m+1\right)}}{\tilde{v}_{\left(m+1\right)\left(m+1\dots N\right)}}\,\frac{\tilde{v}_{\left(m+1\right)m}}{\tilde{v}_{N\left(N-1\right)}\dots\tilde{v}_{32}\tilde{v}_{21}}\\
=-g^{N}\frac{1}{\tilde{v}_{N\left(N-1\right)}\dots\tilde{v}_{32}\tilde{v}_{21}}\,\frac{1}{\tilde{v}_{1\left(1\dots N\right)}}\left[\sum_{m=1}^{N-1}\frac{\tilde{v}_{\left(m+1\dots N\right)\left(m+1\right)}}{\tilde{v}_{\left(m+1\right)\left(m+1\dots N\right)}}\tilde{v}_{\left(m+1\right)m}\right]\\
=-g^{N}\frac{1}{\tilde{v}_{N\left(N-1\right)}\dots\tilde{v}_{32}\tilde{v}_{21}}\,\frac{\tilde{v}_{\left(1\dots N\right)1}}{\tilde{v}_{1\left(1\dots N\right)}},\label{eq:App_Mplus14-1}
\end{multline}
where we have used (\ref{eq:Ident6}) to perform the sum. This indeed
coincides with (\ref{eq:App_Mplus12}).

\section{Derivation of (\ref{eq:Ginvnoninv})}

\label{sec:App_Ginvnoninv}

In order to find the required relation we will use (\ref{eq:Gaugelink_RecRelAlg-1}).
We will subtract two forms of this equation: one with $N=m$ set everywhere except the Wilson line slope
(i.e. the subscript of $\tilde{\mathcal{M}}$), second, with $N=m$ set literally everywhere.
 We get
\begin{multline}
\tilde{\mathcal{M}}_{\varepsilon_{1\dots N}^{+}}^{\left(-+\dots+\right)}\left(k_{1\dots m}\right)=\tilde{\mathcal{M}}_{\varepsilon_{1\dots m}^{+}}^{\left(-+\dots+\right)}\left(k_{1\dots m}\right)\\
+\sum_{i=2}^{m-1}\left[\frac{\tilde{\mathcal{M}}_{\varepsilon_{1\dots N}^{+}}^{\left(-+\dots+\right)}\left(k_{1\dots i}\right)}{\tilde{v}_{\left(1\dots i\right)\left(1\dots N\right)}}-\frac{\tilde{\mathcal{M}}_{\varepsilon_{1\dots m}^{+}}^{\left(-+\dots+\right)}\left(k_{1\dots i}\right)}{\tilde{v}_{\left(1\dots i\right)\left(1\dots m\right)}}\right]
\overline{\mathcal{M}}^{\left(+\rightarrow+\dots+\right)}\left(k_{i+1,\dots,m}\right)\label{eq:App_Ginvnoninv1}
\end{multline}
for $N\geq m$. Using this equation recursively to express the r.h.s
entirely in terms of $\tilde{\mathcal{M}}_{\varepsilon_{1\dots a}^{+}}\left(k_{1\dots a}\right)$
we get
\begin{multline}
\tilde{\mathcal{M}}_{\varepsilon_{1\dots N}^{+}}^{\left(-+\dots+\right)}\left(k_{1\dots m}\right)=\tilde{\mathcal{M}}_{\varepsilon_{1\dots m}^{+}}^{\left(-+\dots+\right)}\left(k_{1\dots m}\right)\\
+\sum_{i=2}^{m-1}\tilde{\mathcal{M}}_{\varepsilon_{1\dots i}^{+}}^{\left(-+\dots+\right)}\left(k_{1\dots i}\right)\,\overline{\mathcal{M}}^{\left(+\rightarrow+\dots+\right)}\left(k_{i+1,\dots,m}\right)\left(\frac{1}{\tilde{v}_{\left(1\dots i\right)\left(1\dots N\right)}}-\frac{1}{\tilde{v}_{\left(1\dots i\right)\left(1\dots m\right)}}\right)\\
\qquad+\sum_{i=2}^{m-1}\sum_{j=2}^{i-1}\tilde{\mathcal{M}}_{\varepsilon_{1\dots j}^{+}}^{\left(-+\dots+\right)}\left(k_{1\dots j}\right)\overline{\mathcal{M}}^{\left(+\rightarrow+\dots+\right)}\left(k_{j+1,\dots,i}\right)\overline{\mathcal{M}}^{\left(+\rightarrow+\dots+\right)}\left(k_{i+1,\dots,m}\right)\\
\Bigg[\frac{1}{\tilde{v}_{\left(1\dots i\right)\left(1\dots N\right)}}\left(\frac{1}{\tilde{v}_{\left(1\dots j\right)\left(1\dots N\right)}}-\frac{1}{\tilde{v}_{\left(1\dots j\right)\left(1\dots i\right)}}\right)-\frac{1}{\tilde{v}_{\left(1\dots i\right)\left(1\dots m\right)}}\left(\frac{1}{\tilde{v}_{\left(1\dots j\right)\left(1\dots m\right)}}-\frac{1}{\tilde{v}_{\left(1\dots j\right)\left(1\dots i\right)}}\right)\Bigg]\\
+\dots\label{App_Ginvnoninv2}
\end{multline}
It can be readily generalized to
\begin{multline}
\tilde{\mathcal{M}}_{\varepsilon_{1\dots N}^{+}}^{\left(-+\dots+\right)}\left(k_{1\dots m}\right)=\tilde{\mathcal{M}}_{\varepsilon_{1\dots m}^{+}}^{\left(-+\dots+\right)}\left(k_{1\dots m}\right)\\
+\sum_{p=1}^{m-2}\sum_{i_{1}=2}^{m-1}\sum_{i_{2}=2}^{i_{1}-1}\dots\sum_{i_{p}=2}^{i_{p-1}-1}\tilde{\mathcal{M}}_{\varepsilon_{1\dots i_{p}}^{+}}^{\left(-+\dots+\right)}\left(k_{1\dots i_{p}}\right)\\
\qquad\qquad\qquad\qquad\qquad\overline{\mathcal{M}}^{\left(+\rightarrow+\dots+\right)}\left(k_{i_{p}+1,\dots,i_{p-1}}\right)\dots\overline{\mathcal{M}}^{\left(+\rightarrow+\dots+\right)}\left(k_{i_{1}+1,\dots,m}\right)\\
\left[a_{i_{1}\dots i_{p}}^{\left(p\right)}\left(N\right)-a_{i_{1}\dots i_{p}}^{\left(p\right)}\left(m\right)\right],\label{eq:App_Ginvnoninv2a}
\end{multline}
where
\begin{equation}
a_{i_{p}\dots i_{p+n-1}}^{\left(n\right)}\left(l\right)=\frac{1}{\tilde{v}_{\left(1\dots i_{p}\right)\left(1\dots l\right)}}\, a_{i_{p+1}\dots i_{p+n-1}}^{\left(n-1\right)}\left(l\right)-\frac{1}{\tilde{v}_{\left(1\dots i_{p}\right)\left(1\dots i_{p-1}\right)}}\, a_{i_{p+1}\dots i_{p+n-1}}^{\left(n-1\right)}\left(i_{p-1}\right)\label{eq:App_Ginvnoninv2d}
\end{equation}
with the initial conditions
\begin{equation}
a^{\left(0\right)}\left(l\right)=1,\,\,\,\,\,\,\, a_{i_{p}\dots i_{p+n-1}}^{\left(n\right)}\left(i_{q}\right)=0\,\,\,\,\,\textrm{for}\,\, q<1.
\end{equation}
This binary tree can be simplified by an extensive use of the identities
from Appendix \ref{sec:Identities}, notably equation (\ref{eq:Ident3}).
For example, for $p=1$ in the sum in (\ref{eq:App_Ginvnoninv2a})
we have
\begin{multline}
a_{i_{1}}^{\left(1\right)}\left(N\right)-a_{i_{1}}^{\left(1\right)}\left(m\right)=\frac{1}{\tilde{v}_{\left(1\dots i_{1}\right)\left(1\dots N\right)}}-\frac{1}{\tilde{v}_{\left(1\dots i_{1}\right)\left(1\dots m\right)}}\\
=\frac{z_{1\dots i_{1}}}{z_{1\dots N}}\frac{\tilde{v}_{\left(1\dots N\right)\left(1\dots m\right)}}{\tilde{v}_{\left(1\dots i_{1}\right)\left(1\dots N\right)}\tilde{v}_{\left(1\dots i_{1}\right)\left(1\dots m\right)}}=\frac{\tilde{v}_{\left(1\dots m\right)\left(1\dots N\right)}}{\tilde{v}_{\left(1\dots i_{1}\right)\left(1\dots N\right)}\tilde{v}_{\left(1\dots m\right)\left(1\dots i_{1}\right)}}.\label{App_Ginvnoninv2c}
\end{multline}
Next, for $p=2$ we have
\begin{multline}
a_{i_{1}i_{2}}^{\left(2\right)}\left(N\right)-a_{i_{1}i_{2}}^{\left(2\right)}\left(m\right)=\frac{1}{\tilde{v}_{\left(1\dots i_{1}\right)\left(1\dots N\right)}}\, a_{i_{2}}^{\left(1\right)}\left(N\right)-\frac{1}{\tilde{v}_{\left(1\dots i_{1}\right)\left(1\dots m\right)}}\, a_{i_{2}}^{\left(1\right)}\left(i_{1}\right)\\
=\frac{z_{1\dots i_{2}}^{2}}{z_{1\dots i_{1}}z_{1\dots N}}\frac{-\tilde{v}_{\left(1\dots N\right)\left(1\dots m\right)}}{\tilde{v}_{\left(1\dots i_{1}\right)\left(1\dots N\right)}\tilde{v}_{\left(1\dots i_{2}\right)\left(1\dots m\right)}\tilde{v}_{\left(1\dots i_{2}\right)\left(1\dots i_{1}\right)}}=\frac{\tilde{v}_{\left(1\dots m\right)\left(1\dots N\right)}}{\tilde{v}_{\left(1\dots i_{1}\right)\left(1\dots N\right)}\tilde{v}_{\left(1\dots m\right)\left(1\dots i_{2}\right)}\tilde{v}_{\left(1\dots i_{1}\right)\left(1\dots i_{2}\right)}}
\end{multline}
and so on. In fact, the solution to (\ref{eq:App_Ginvnoninv2d}) reads
\begin{equation}
a_{i_{p}\dots i_{p+n-1}}^{\left(n\right)}\left(l\right)=\frac{\tilde{v}_{\left(1\dots i_{p-1}\right)\left(1\dots l\right)}}{\tilde{v}_{\left(1\dots i_{p+n-1}\right)\left(1\dots l\right)}}\frac{1}{\tilde{v}_{\left(1\dots i_{p-1}\right)\left(1\dots i_{p+n-1}\right)}\tilde{v}_{\left(1\dots i_{p}\right)\left(1\dots i_{p+n-1}\right)}\dots\tilde{v}_{\left(1\dots i_{p+n-2}\right)\left(1\dots i_{p+n-1}\right)}},\label{eq:App_Ginvnoninv_4}
\end{equation}
for $p>1$, as can be easily verified. Therefore, the equation (\ref{eq:App_Ginvnoninv2a})
can be written as
\begin{multline}
\tilde{\mathcal{M}}_{\varepsilon_{1\dots N}^{+}}^{\left(-+\dots+\right)}\left(k_{1\dots m}\right)=\tilde{\mathcal{M}}_{\varepsilon_{1\dots m}^{+}}^{\left(-+\dots+\right)}\left(k_{1\dots m}\right)\\
+\sum_{p=1}^{m-2}\sum_{i_{1}=2}^{m-1}\sum_{i_{2}=2}^{i_{1}-1}\dots\sum_{i_{p}=2}^{i_{p-1}-1}\tilde{\mathcal{M}}_{\varepsilon_{1\dots i_{p}}^{+}}^{\left(-+\dots+\right)}\left(k_{1\dots i_{p}}\right)\\
\qquad\qquad\qquad\qquad\qquad\qquad\overline{\mathcal{M}}^{\left(+\rightarrow+\dots+\right)}\left(k_{i_{p}+1,\dots,i_{p-1}}\right)\dots\overline{\mathcal{M}}^{\left(+\rightarrow+\dots+\right)}\left(k_{i_{1}+1,\dots,m}\right)\\
\frac{\tilde{v}_{\left(1\dots m\right)\left(1\dots N\right)}}{\tilde{v}_{\left(1\dots i_{p}\right)\left(1\dots N\right)}\tilde{v}_{\left(1\dots m\right)\left(1\dots i_{p}\right)}}\,\frac{1}{\tilde{v}_{\left(1\dots i_{1}\right)\left(1\dots i_{p}\right)}\dots\tilde{v}_{\left(1\dots i_{p-1}\right)\left(1\dots i_{p}\right)}}.\label{App_Ginvnoninv3}
\end{multline}

\section{Proof of recursion (\ref{eq:Gaugelink_RecRelAlg-2})}

\label{sec:App_RecProof}

First, we use (\ref{App_Ginvnoninv3}) to rewrite (\ref{eq:Gaugelink_RecRelAlg-1})
purely in terms of gauge invariant amplitudes. The resulting equation
is, however, very complicated. In order to simplify it we note the
following identity
\begin{multline}
\overline{\mathcal{M}}^{\left(+\rightarrow+\dots+\right)}\left(k_{i_{p}+1,\dots,i_{p-1}}\right)\dots\,\,\overline{\mathcal{M}}^{\left(+\rightarrow+\dots+\right)}\left(k_{i_{1}+1,\dots,m}\right)\\
=\frac{z_{i_{p}+1\dots i_{p-1}}\dots z_{i_{2}+1\dots i_{1}}}{z_{i_{p}+1}\dots z_{i_{1}+1}}\,\frac{z_{i_{1}+1\dots m} z_{i_{p}+1}}{z_{i_{p}+1\dots m}}\,\tilde{v}_{\left(i_{p-1}+1\right)i_{p-1}}\dots\tilde{v}_{\left(i_{1}+1\right)i_{1}}\\
\overline{\mathcal{M}}^{\left(+\rightarrow+\dots+\right)}\left(k_{i_{p}+1,\dots,m}\right).\label{eq:Mbarprod}
\end{multline}
It follows directly from (\ref{eq:Mbar+}). Using this we can write
(\ref{eq:Gaugelink_RecRelAlg-1}) as 
\begin{multline}
\tilde{\mathcal{M}}_{\varepsilon_{1\dots N}^{+}}^{\left(-+\dots+\right)}\left(k_{1\dots N}\right)=\overline{\mathcal{M}}^{\left(+\rightarrow-\dots+\right)}\left(k_{1,\dots,N}\right)\\
+\sum_{m=2}^{N-1}\tilde{\mathcal{M}}_{\varepsilon_{1\dots m}^{+}}^{\left(-+\dots+\right)}\left(k_{1\dots m}\right)\frac{1}{\tilde{v}_{\left(1\dots m\right)\left(1\dots N\right)}}\overline{\mathcal{M}}^{\left(+\rightarrow+\dots+\right)}\left(k_{m+1,\dots,N}\right)\\
+\sum_{m=2}^{N-1}\sum_{p=1}^{m-2}\sum_{i_{1}=2}^{m-1}\sum_{i_{2}=2}^{i_{1}-1}\dots\sum_{i_{p}=2}^{i_{p-1}-1}\tilde{\mathcal{M}}_{\varepsilon_{1\dots i_{p}}^{+}}^{\left(-+\dots+\right)}\left(k_{1\dots i_{p}}\right)\overline{\mathcal{M}}^{\left(+\rightarrow+\dots+\right)}\left(k_{i_{p}+1,\dots,N}\right)\\
\frac{z_{i_{p}+1\dots i_{p-1}}\dots z_{i_{2}+1\dots i_{1}}}{z_{i_{p-1}+1}\dots z_{i_{1}+1}}\,\frac{z_{i_{1}+1\dots m}}{z_{i_{p}+1\dots N}}\frac{z_{m+1\dots N}}{z_{m+1}}\,\\
\frac{\tilde{v}_{\left(m+1\right)m}}{\tilde{v}_{\left(1\dots i_{p}\right)\left(1\dots N\right)}\tilde{v}_{\left(1\dots m\right)\left(1\dots i_{p}\right)}}\frac{\tilde{v}_{\left(i_{1}+1\right)i_{1}}\dots\tilde{v}_{\left(i_{p-1}+1\right)i_{p-1}}}{\tilde{v}_{\left(1\dots i_{1}\right)\left(1\dots i_{p}\right)}\dots\tilde{v}_{\left(1\dots i_{p-1}\right)\left(1\dots i_{p}\right)}}.\label{eq:App_RecRel1}
\end{multline}
The amplitudes on the r.h.s. depend only on single summation variable
$i_{p}$, and it turns out that one can perform the remaining sums.
To this end, let us reorganize the sums as follows
\begin{equation}
\sum_{m=2}^{N-1}\sum_{p=1}^{m-2}\sum_{i_{1}=2}^{m-1}\sum_{i_{2}=2}^{i_{1}-1}\dots\sum_{i_{p}=2}^{i_{p-1}-1}=\sum_{p=1}^{N-3}\sum_{i_{p}=2}^{N-1}\sum_{i_{p-1}=i_{p}+1}^{N-1}\dots\sum_{i_{1}=i_{2}+1}^{N-1}\sum_{m=i_{1}+1}^{N-1}.
\end{equation}
Let us also rename variables $m\rightarrow i_{0}$, $i_{p}\rightarrow m$.
This allows to rewrite (\ref{eq:App_RecRel1}) as 
\begin{multline}
\tilde{\mathcal{M}}_{\varepsilon_{1\dots N}^{+}}^{\left(-+\dots+\right)}\left(k_{1\dots N}\right)=\overline{\mathcal{M}}^{\left(+\rightarrow-\dots+\right)}\left(k_{1,\dots,N}\right)\\
+\sum_{m=2}^{N-1}\tilde{\mathcal{M}}_{\varepsilon_{1\dots m}^{+}}^{\left(-+\dots+\right)}\left(k_{1\dots m}\right)\overline{\mathcal{M}}^{\left(+\rightarrow+\dots+\right)}\left(k_{m+1,\dots,N}\right)\frac{1}{z_{m+1\dots N}\tilde{v}_{\left(1\dots m\right)\left(1\dots N\right)}}\\
\Bigg\{ z_{m+1\dots N}+\sum_{p=0}^{N-4}\sum_{i_{p}=m+1}^{N-1}\dots\sum_{i_{0}=i_{1}+1}^{N-1}\frac{z_{m+1\dots i_{p}}\dots z_{i_{1}+1\dots i_{0}}z_{i_{0}+1\dots N}}{z_{i_{p}+1}\dots z_{i_{0}+1}}\\
\frac{\tilde{v}_{\left(i_{0}+1\right)i_{0}}\dots\tilde{v}_{\left(i_{p}+1\right)i_{p}}}{\tilde{v}_{\left(1\dots i_{0}\right)\left(1\dots m\right)}\dots\tilde{v}_{\left(1\dots i_{p}\right)\left(1\dots m\right)}}\Bigg\}.\label{eq:App_RecRel2}
\end{multline}
The tower of sums above can be utilized as follows. First, let us
introduce

\begin{multline}
\kappa_{mN}=z_{m+1\dots N}\\
+\sum_{p=0}^{N-4}\sum_{i_{p}=m+1}^{N-1}\dots\sum_{i_{0}=i_{1}+1}^{N-1}\frac{z_{m+1\dots i_{p}}\dots z_{i_{1}+1\dots i_{0}}z_{i_{0}+1\dots N}}{z_{i_{p}+1} \dots z_{i_{0}+1}}\frac{\tilde{v}_{\left(i_{0}+1\right)i_{0}}\dots\tilde{v}_{\left(i_{p}+1\right)i_{p}}}{\tilde{v}_{\left(1\dots i_{0}\right)\left(1\dots m\right)}\dots\tilde{v}_{\left(1\dots i_{p}\right)\left(1\dots m\right)}}\label{eq:App_kappanm}
\end{multline}
It can be rewritten as
\begin{equation}
\kappa_{mN}=\tilde{z}_{mN}+\sum_{p=0}^{N-4}\sum_{i_{p}}\dots\sum_{i_{0}}\,\tilde{z}_{mi_{p}}h_{i_{p}}\left(m\right)\tilde{z}_{i_{p}i_{p-1}}\dots h_{i_{0}}\left(m\right)\tilde{z}_{i_{0}N},\label{eq:App_kappanm1}
\end{equation}
where
\begin{equation}
\tilde{z}_{ij}=\Theta\left(j-i-1\right)z_{\left(i+1\right)\dots j}\label{eq:App_ztild}
\end{equation}
with $\Theta\left(j-i\right)$ being the Heaviside step function,
and
\begin{equation}
h_{i}\left(m\right)=\frac{\tilde{v}_{\left(i+1\right)i}}{z_{i+1} \tilde{v}_{\left(1\dots i\right)\left(1\dots m\right)}}.\label{eq:App_hidef}
\end{equation}
We can consider $\tilde{z}_{ij}$ as a ``free propagator'' and $h_{i}\left(m\right)$
as a ``vertex'', and (\ref{eq:App_kappanm1}) as the equation for
the ``full propagator''. Graphically it can be represented as

\begin{tabular}{>{\centering}m{0.87\columnwidth}>{\centering}m{0.05\columnwidth}}
\bigskip{}

\centering{}\includegraphics[width=0.87\textwidth]{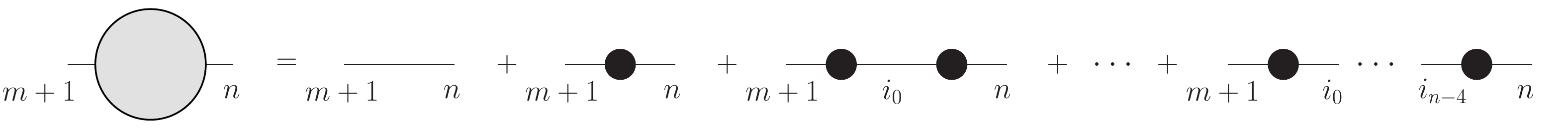} & \centering{}(\myref)\label{eq:App_prop1}\tabularnewline
\end{tabular}

\noindent where the blob represents $\kappa_{mn}$, the black dots
represent vertices $h_{i}$, and the lines represent propagators $\tilde{z}_{ij}$.
At each vertex there is a summation over the corresponding index.
It is easy to see that (\ref{eq:App_kappanm1}) has the following
factorization property, graphically

\begin{tabular}{>{\centering}m{0.87\columnwidth}>{\centering}m{0.05\columnwidth}}
\bigskip{}

\centering{}\includegraphics[width=0.55\textwidth]{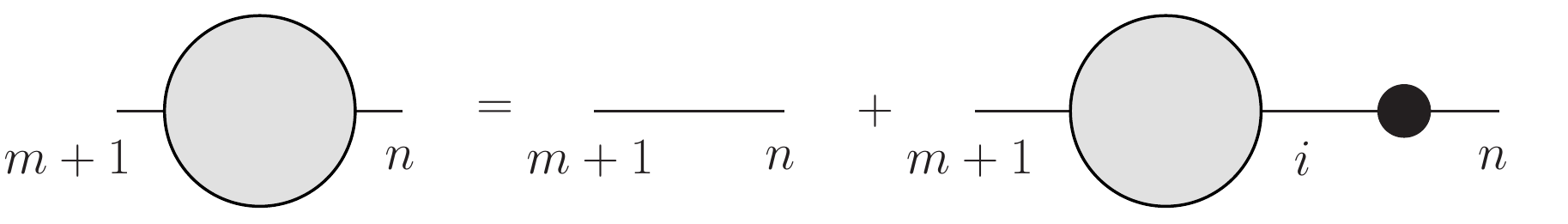} & \centering{}(\myref)\label{eq:App_prop2}\tabularnewline
\end{tabular}

\noindent or
\begin{equation}
\kappa_{mn}=\tilde{z}_{mn}+\sum_{i}\kappa_{mi}h_{i}\left(m\right)\tilde{z}_{\left(i+1\right)n}.\label{eq:App_prop3}
\end{equation}
We will prove, that the solution to this equation reads
\begin{equation}
\kappa_{mn}=\frac{z_{1\dots n}\tilde{v}_{\left(1\dots m\right)\left(1\dots n\right)}}{\tilde{v}_{\left(1\dots m+1\right)\left(m+1\right)}}.\label{eq:App_Kmnsol}
\end{equation}
First, consider the sum on the r.h.s. of (\ref{eq:App_prop3}) inserting
the above ansatz
\begin{multline}
\sum_{i}\Theta\left(n-i-1\right)\Theta\left(i-m-1\right)\frac{z_{1\dots i}\tilde{v}_{\left(1\dots m\right)\left(1\dots i\right)}}{\tilde{v}_{\left(1\dots m+1\right)\left(m+1\right)}}\,\frac{\tilde{v}_{\left(i+1\right)i}\, z_{i+1\dots n}}{z_{i+1} \tilde{v}_{\left(1\dots i\right)\left(1\dots m\right)}}\\
=\frac{z_{1\dots m}}{\tilde{v}_{\left(1\dots m+1\right)\left(m+1\right)}}\sum_{i=m+1}^{n-1}\frac{\tilde{v}_{\left(i+1\dots n\right)\left(i+1\right)}}{\tilde{v}_{\left(i+1\right)\left(i+1\dots n\right)}}\,\tilde{v}_{\left(i+1\right)i}=-\frac{z_{1\dots m}\tilde{v}_{\left(m+1\dots n\right)\left(m+1\right)}}{\tilde{v}_{\left(1\dots m+1\right)\left(m+1\right)}}\label{eq:App_Kmnsol1}
\end{multline}
thanks to the identity (\ref{eq:Ident6}). The complete r.h.s. of
(\ref{eq:App_prop3}) now reads
\begin{multline}
z_{\left(m+1\right)\dots n}-\frac{z_{1\dots m}\tilde{v}_{\left(m+1\dots n\right)\left(m+1\right)}}{\tilde{v}_{\left(1\dots m+1\right)\left(m+1\right)}}=z_{\left(m+1\right)\dots n}-\frac{z_{m+1\dots n}\tilde{v}_{\left(m+1\right)\left(m+1\dots n\right)}}{\tilde{v}_{\left(m+1\right)\left(1\dots m\right)}}\\
=\frac{z_{\left(m+1\right)\dots n}}{\tilde{v}_{\left(m+1\right)\left(1\dots m\right)}}\left[\tilde{v}_{\left(m+1\right)\left(1\dots m\right)}-\tilde{v}_{\left(m+1\right)\left(m+1\dots n\right)}\right]=\frac{z_{m+1}\tilde{v}_{\left(1\dots n\right)\left(1\dots m\right)}}{\tilde{v}_{\left(m+1\right)\left(1\dots m\right)}}=\frac{z_{1\dots n}\tilde{v}_{\left(1\dots m\right)\left(1\dots n\right)}}{\tilde{v}_{\left(1\dots m+1\right)\left(m+1\right)}}\label{eq:App_Kmnsol2}
\end{multline}
where we have used identities from Appendix \ref{sec:Identities}.
We see that this is the same as (\ref{eq:App_Kmnsol}), thus we have
accomplished the proof.

It is now easy to read out the expression for $K_{mN}$ defined in (\ref{eq:Gaugelink_RecRelAlg-2}).
It follows from comparison of (\ref{eq:Gaugelink_RecRelAlg-2}) and
(\ref{eq:App_RecRel2}) and simply reads
\begin{equation}
K_{mN}=\frac{z_{1\dots N}}{z_{m+1\dots N}\tilde{v}_{\left(1\dots m+1\right)\left(m+1\right)}}.\label{eq:App_KmN}
\end{equation}

\bibliographystyle{ieeetr}
\bibliography{small_x,LCPT}

\end{document}